\documentclass[%
 aip,
 amsmath,amssymb,
 reprint,%
]{revtex4-2}

\usepackage{graphicx}
\usepackage{dcolumn}
\usepackage{bm}

\usepackage[utf8]{inputenc}
\usepackage[T1]{fontenc}
\usepackage{mathptmx}

\usepackage{algorithmic,algorithm}
\usepackage{subcaption}
\usepackage{xcolor}
\usepackage{amssymb}
\usepackage{amsmath}
\usepackage{amsfonts}
\usepackage{comment}
\usepackage{graphicx}
\usepackage{newtxtext,newtxmath}
\usepackage{tikz}
\usepackage{pgfplots}
\usepackage{float}
\usepackage{placeins}
\usetikzlibrary{external}
\newcommand{\tens}[1] {\bm{\mathsf{#1}}}

\begin{document}

\preprint{AIP/123-QED}

\title[Tensor-train approximation of the chemical master equation and its application for parameter inference]{Tensor-train approximation of the chemical master equation and its application for parameter inference}

\author{Ion Gabriel Ion}
\affiliation{Centre for Computational Engineering, Technische Universit\"at Darmstadt}
\affiliation{Department of Electrical Engineering, Technische Universit\"at Darmstadt}
 \email{ion@temf.tu-darmstadt.de}
\author{Christian Wildner}%
\affiliation{Department of Electrical Engineering, Technische Universit\"at Darmstadt}

\author{Dimitrios Loukrezis}
\affiliation{Centre for Computational Engineering, Technische Universit\"at Darmstadt}
\affiliation{Department of Electrical Engineering, Technische Universit\"at Darmstadt}

\author{Heinz Koeppl}%
\affiliation{Centre for Computational Engineering, Technische Universit\"at Darmstadt}
\affiliation{ Centre for Synthetic Biology, Technische Universit\"at Darmstadt}
\affiliation{Department of Electrical Engineering, Technische Universit\"at Darmstadt}

\author{Herbert De Gersem}%
\affiliation{Centre for Computational Engineering, Technische Universit\"at Darmstadt}
\affiliation{Department of Electrical Engineering, Technische Universit\"at Darmstadt}


\begin{abstract}
In this work, we perform Bayesian inference tasks for the chemical master equation in the tensor-train format. 
The tensor-train approximation has been proven to be very efficient in representing high dimensional data arising from the explicit representation of the chemical master equation solution. 
An additional advantage of representing the probability mass function in the tensor train format is that parametric dependency can be easily incorporated by introducing a tensor product basis expansion in the parameter space. 
Time is treated as an additional dimension of the tensor and a linear system is derived to solve the chemical master equation in time.
We exemplify the tensor-train method by performing inference tasks such as smoothing and parameter inference using the tensor-train framework. A very high compression ratio is observed for storing the probability mass function of the solution. Since all linear algebra operations are performed in the tensor-train format, a significant reduction of the computational time is observed as well. \end{abstract}

\maketitle

\section{Introduction}
Traditional chemical kinetic models use ordinary differential equations (ODEs) to predict the concentrations of the involved molecule types.
The evolution of the corresponding probability distribution is given by the chemical master equation (CME) \cite{gillespie1992rigorous} which, in principle, can be solved by numerical integration. In practice, the state space even of simple models is too large for a naive integration of the CME. Therefore, a number of approximation techniques have been developed over the years, e.g. stochastic simulation methods \cite{gibson2000efficient, hemberg2007perfect, gillespie_ssa} and suitable time and space discretizations \cite{engblom2009spectral, deuflhard2008adaptive}.
Many CME approximations are based on the observation that the probability mass is often concentrated on a small fraction of the state space.  
For example, the finite state projection method solves the CME on a rectangular subspace with appropriate boundary conditions \cite{munsky_2006, burrage_2006}. 
A problem here is that the region where the significant part of the  probability mass function (PMF) is located may change over time. 
This is tackled in the sliding window method, where the region of interest is adapted based on the solution at the previous time step \cite{wolf_2010}. 
Unfortunately, the computational cost of these methods grows exponentially with the number of species, due to the fact that the system states must be labeled explicitly to cast the CME into a ODE. 

A different line of research has explored approximations of the CME based on low-rank tensor formats \cite{cme_time, kazeev_2014, dinh2020adaptive, vo2017adaptive, jahnke_2008}. 
The idea is to project the probability distribution onto a subspace of the tensor-product space induced by the reaction system. 
The solution is then propagated by a small time step and projected back onto the chosen space. 
Alternatively, considering time as an additional dimension, a joint approximation of the space-time system in a low-rank tensor format can be obtained \cite{cme_time, kazeev_2014}. 
\color{black}
The low-rank tensor representation not only preserves the structure of the CME, but is also much more memory-efficient compared to the matrix representation approach\cite{gupta2017}. 
In addition, the low-rank tensor representation allows for accurate, dynamic approximations via rank rounding. 

Low-rank tensor decompositions have also been considered in the context of parameter-dependent CMEs, however, with limited applications so far\cite{cme_time}.
For example, considering systems and synthetic biology,  stochastic chemical kinetics are used to construct quantitatively predictive models of biomolecular circuits. This requires the solution of an inverse problem, where it is often necessary to estimate the rate parameters of a structurally known candidate system from time course data. For population snapshot data, e.g. obtained from flow cytometry measurements, calibration via moment-based inference is a well-established method\cite{kuegler_2012, zechner_2012, froehlich_2016}. In the last decade, advances in fluorescence microscopy have led to an increasing availability of single-cell time course data. When the number of observed cells is small, standard moment-based methods break down because they rely on the central limit theorem to compute a likelihood function for sample moments \cite{munsky_2018, cao_2019, Bronstein2018AVA}. In such a scenario, inference based on the path likelihood of individual trajectories provides a principled way to extract more information from the data. Unfortunately, these approaches are computationally challenging since they require integrating the CME multiple times for different parameter configurations. More effective approaches can be obtained by approximating the likelihood, e.g. by Gaussian moment closure\cite{milner_2013} or the linear noise approximation\cite{stathopoulos_2013,fearnhead_2014}. While reducing the computational demand significantly, the underlying approximations are not applicable to systems with low copy number of species, such as single genes. Alternatively, approximate likelihoods can be computed via particle filtering \cite{golightly_2011}. For Bayesian parameter inference, the approximate likelihood expression are typically used within a Markov chain Monte Carlo scheme. Alternatively, the parameters can be included into an augmented state space allowing for direct estimation via sequential Monte Carlo\cite{zechner_2014}.

In this work, we suggest a framework for performing Bayesian inference tasks for the parameter-dependent CME, by exploiting the so-called tensor-train (TT) decomposition \cite{oseledets1} to approximate the joint distribution over the CME states and parameters. 
For that purpose, we construct an explicit representation of the evolution operator in the TT format and show that it can be constructed without ever assembling the corresponding matrix.
The TT format has the advantage that the storage requirement scales linearly with respect to the number of dimensions, while at the same time being a numerically robust tensor decomposition \cite{Kolda2009TensorDA,oseledets1}.
We also develop a time-domain solver based on the time-dependent alternating minimal energy (tAMEn) algorithm\cite{dolgov_tamen}, which additionally incorporates parameter dependence.
To that end, we combine the the state space and the parameter space into a higher-dimensional tensor-product space. 
The parameter dependence is expressed by means of a B-spline basis expansion and a Galerkin formulation is employed to derive the multilinear system with respect to the full tensor.
Since typically every reaction is governed by an individual rate constant, the parameters can be seamlessly included in the tensor representation, thus allowing for efficiently solving the joint system. 
In practice, however, the system parameters are often unknown. 
Therefore, we develop a framework for filtering, smoothing, and parameter inference based on the efficient TT representation of the joint system. 
The proposed framework allows us to perform Bayesian inference for the model parameters with a single forward-backward pass, as demonstrated on several synthetic examples.

The remaining of this paper is organized as follows. 
In section~\ref{sec:tt-cme}, we recall the CME and explain how it can be expressed in tensor format. 
Next, in section~\ref{sec:solve-tt-cme}, we present the TT decomposition, apply it to the tensor-formatted CME, and present a TT-based solution method.
In section~\ref{sec:infer-tt-cme}, we consider the setting of a CME with parameter dependencies, which we include in the TT-based CME format.
Subsequently, we exploit the TT-format of the parameter-dependent CME for inference tasks, such as filtering, smoothing, and parameter identification.
Numerical results are presented in section~\ref{sec:num-results}, where we validate the TT-based CME solver and showcase the benefits of performing inference in the TT-format.
The paper closes with our conclusions, presented in section~\ref{sec:conclusion}.

\section{Chemical Master Equation in Tensor Format}
\label{sec:tt-cme}
We consider a well-mixed reaction system with $d$ chemical species denoted with $\left\{S_1,...,S_d\right\}$ involved in $M$ reactions. 
We consider reactions of type:
\begin{align}
q_{m,1} S_1 +  \cdots + q_{m,d} S_d \longrightarrow s_{m,1} S_1 + \cdots + s_{m,d} S_d, \label{eq:reaction}
\end{align}
with $m=1,\dots,M$ and $q_{m,k},s_{m,k} \in \mathbb{N}_0$, $k=1,\dots,d$.
The state of the system at time $t$ is described by the vector $\bm{x} \in \mathbb{N}_0^{d}$ which contains the number of elements per species at that time instant.
To describe the change in the state vector after reaction $m$ occurs, we introduce the stoichiometric change vector $\bm{\nu}^{(m)} \in \mathbb{Z}^d$, the $k$th element of which is given as $\nu_k^{(m)} = s_{m,k} - q_{m,k}$.
The change in the state vector due to the $m$th reaction  is then given as $\bm{x} \rightarrow \bm{x} +\bm{\nu}^{(m)}$.

Assuming a stochastic model of the system, the state vector $\bm{x}$ is a realization of a continuous-time jump process $\left\{\bm{X}\left(t\right)\right\}_{t\geq 0}$.
Then, the evolution of the time-dependent PMF 
\begin{align}
p(t, \bm{x}) = p_t(\bm{x}) &= \text{Pr}\left( \bm{X}\left( t \right) = \bm{x} \right) \nonumber \\
&= \text{Pr}\left( X_1\left(t\right)=x_1, \dots, X_d\left(t\right)=x_d\right)
\label{pmf}
\end{align}
is described by the so-called chemical master equation (CME) \cite{gillespie1992rigorous}, such that
\begin{subequations}
	\begin{align}
	\frac{{\rm d} p_t(\bm{x})}{{\rm d} t} &= \sum_{m=1}^M \left\{ \alpha_m({\bm{x}-\bm{\nu}^{(m)}}) p_t({\bm{x}-\bm{\nu}^{(m)}}) - \alpha_m\left(\bm{x}\right) p_t\left(\bm{x}\right)\right\} \label{eq:cme}, \\ 
	p_0(\bm{x}) &= {P}^{(0)}({\bm{x}}),  \label{eq:ivp}
	\end{align}
\end{subequations} 
where ${P}^{( 0)}$ is the initial probability and $\alpha_m$ is called the propensity function. 
For a well-mixed system at thermal equilibrium, the mass-action propensity reads
\begin{align}
\alpha_m(\bm{x}) = c_m \prod\limits_{k=1}^{d} \frac{x_k!}{q_{m,k}!(x_k - q_{m,k})!}, \label{eq:props}
\end{align}
where $c_m$ is the so-called specific probability rate, which is a measure for the probability that reaction $m$ occurs.

Until now, the CME has been defined for an infinite state space $\mathbb{N}_0^{d}$, which is computationally intractable, thus inappropriate for a numerical solution.
To that end, a truncation of the state space is necessary, such that $x_k < n_k$, $k=1,\dots,d$.
We denote the truncated state space as $\mathcal{X} = \left\{\bm{x} \in \mathbb{N}_0^d \,|\, x_k < n_k, k=1,\dots,d\right\}$. The choice of a box domain truncation is not strictly necessary, however, it is beneficial for the TT format used in the following.
All states in $\mathcal{X}$ can be uniquely indexed as $\bm{x}{(\bm{i})}$, where $\bm{i} = \left(i_1,\dots,i_d\right) \in \mathbb{N}^d$ and $x_k{(i_k)} = i_k - 1$.
Accordingly, $i_k = 1, \dots, n_k$.

Using the truncated state space defined above and for a given time instance $t$, the PMF $p\left(t, \bm{x}\right)$ can be represented as a multidimensional array $\tens{p}\left(t\right) \in \mathbb{R}^{n_1 \times \cdots \times n_d}$, the elements of which are given as 
\begin{equation}
\label{eq:pmf-tensor}
\tens{p}_{\bm{i}}(t) = p(t, \bm{x}{(\bm{i})} ).
\end{equation}
In the following, we shall refer to such multidimensional arrays as \emph{tensors} \cite{Kolda2009TensorDA}.
The evolution equation \eqref{eq:cme} can then be written in tensor format, such that
\begin{align}
\frac{{\rm d} \tens{p}(t)}{{\rm d} t} = \tens{A} \tens{p}(t),
\end{align}
where $\tens{A} \in {\mathbb{R}^{\left(n_1 \times \cdots \times n_d\right) \times \left(n_1 \times \cdots \times n_d\right)}}$ is a tensor-operator, also called a tensor-matrix, that acts on the tensor $\tens{p}(t)$.
Tensor-operators can be seen as generalizations of the commonly employed matrix-based operators to more than two dimensions.
The elements of the CME tensor-operator are given as
\begin{equation} 
\label{eq:cme_op}
\tens{A}_{\bm{i, j}} = \sum_{m=1}^{M}  \alpha_m(\bm{x}{(\bm{i})} - \bm{\nu}^{(m)}) \delta_{\bm{x}{(\bm{i})} - \bm{\nu}^{(m)}}^{\bm{x}{(\bm{j})}} - \alpha_m\left(\bm{x}(\bm{i})\right) \delta_{\bm{x}{(\bm{i})}}^{\bm{x}{(\bm{j})}}, 
\end{equation}
where $\delta_{\bm{i}}^{\bm{j}} = \delta_{i_1}^{j_1} \cdots \delta_{i_d}^{j_d}$, with $\delta_{i_k}^{j_k}$ denoting the Kronecker delta.
Accordingly, the product between a tensor-operator and a tensor, the result of which is elementwise given as 
\begin{equation}
\label{eq:tensor-matrix-vector-product}
\left(\tens{A} \tens{p}\left(t\right)\right)_{\bm{i}} = \sum_{\bm{j}} \tens{A}_{\bm{i,j}} \tens{p}_{\bm{j}}(t),   
\end{equation}
can be seen as a generalization of the standard matrix-vector product.

The complexity for storing for storing the tensor $\tens{p}(t)$, which contains all state probabilities at time instance $t$, is $\mathcal{O}(n^d)$, where $n=\max_k \left\{n_k\right\}$. 
Therefore, even if a truncated state space is employed, the storage needs can become intractable even for a relatively small number of species. 
The exponential dependence of storage needs to the number of species is one manifestation of the so-called curse of dimensionality \cite{oseledets_curse}. 
As a remedy to this problem, low-rank tensor formats \cite{Kolda2009TensorDA} can be employed, such as the TT decomposition \cite{oseledets1} discussed next.

In the following, a commonly employed operation between tensors, tensor-operators, or matrices, is the Kronecker product. 
The Kronecker product between two tensors $\tens{x}\in\mathbb{R}^{n_1\times\cdots\times n_p}$ and $\tens{y}\in\mathbb{R}^{m_1\times\cdots\times m_q}$ is defined elementwise as
\begin{equation}
(\tens{x}\otimes\tens{y})_{\bm{i} \bm{j}} = \tens{x}_{\bm{i}} \tens{y}_{\bm{j}}.
\end{equation}
The definition holds also for matrices and vectors, as they can be interpreted as two-dimensional and one-dimensional tensors, respectively.

\section{Solving the Chemical Master Equation in the Tensor-Train Format}
\label{sec:solve-tt-cme}
 
\subsection{Tensor-train decomposition}
As discussed in the previous section, the size of a tensor scales exponentially with the number of dimensions, equivalently, number of species in this work.
To mitigate the curse of dimensionality, we employ tensor decompositions, resulting in tensor formats whose sizes scale linearly with the number of dimensions, instead of exponentially. 
Several tensor decompositions have been developed over the last decades, resulting in better-scaling tensor formats \cite{Kolda2009TensorDA}. 
In this work, we focus on the so-called tensor-train (TT) decomposition, which combines linear complexity scaling with respect to the dimensions and computational stability \cite{oseledets1}.

A tensor $\tens{x} \in \mathbb{R}^{n_1 \times \cdots \times n_d}$ is said to be in the TT format if it can be elementwise written as
\begin{align}
\label{eq:tt-format}
\tens{x}_{\bm{i}} = 
\sum_{r_1=1}^{R_1} \sum_{r_2=1}^{R_2} \cdots \sum_{r_{d-1}=1}^{R_{d-1}} 
\tens{g}^{(1)}_{1 i_1 r_1} \tens{g}^{(2)}_{r_1 i_2 r_2} \cdots \tens{g}^{(d)}_{r_{d-1} i_d 1},
\end{align}
where the three-dimensional tensors $\tens{g}^{(k)} \in \mathbb{R} ^ {R_{k-1} \times n_k \times R_k}$ are called the TT-cores and $\bm{R} = (1,R_1,...,R_{d-1},1)$ are called the TT-ranks. 
The storage complexity of a tensor in the TT format is reduced to $\mathcal{O}(N R^2 d)$, i.e. it is linear with respect to the number of dimensions $d$. 
Moreover, all basic multilinear algebraic operations scale also linearly with the dimensions and polynomially with the  TT-ranks \cite{oseledets1}.
It should be noted that the TT-ranks grow after a multilinear algebraic operation is performed in the TT-format.
Therefore, a rank-reduction procedure is performed after the operation, called rounding \cite{oseledets1}.
Rounding decreases the TT-rank of the tensor while maintaining a prescribed accuracy $\epsilon$ and its complexity is $\mathcal{O}(R^3 N d)$.

An exact TT decomposition of a full tensor typically leads to high ranks $\bm{R}$, hence, to high storage needs and computational costs as well. 
However, in many cases, using a low-rank TT approximation $\widetilde{\tens{A}} \approx \tens{A}$ is sufficient . 
If the full tensor is available, a low-rank TT approximation can be efficiently computed with $d$ sequential singular value decompositions (SVDs) of auxiliary matrices $\bm{A}^{(k)} \in \mathbb{R}^{n_k \times \left(n_1 n_2 \cdots n_{k-1} n_{k+1} \cdots n_d\right)}$ \cite{oseledets1}.
For tensors defined implicitly by multidimensional functions, e.g. similar to \eqref{eq:pmf-tensor}, efficient interpolation-based TT approximation methods are available \cite{savostyanov2011fast, oseledets2010tt}, in which case the assembly of the full tensor is not necessary.

A tensor-operator $\tens{A} \in \mathbb{R}^{\left(n_1\times \cdots \times n_d\right) \times \left(m_1 \times \cdots \times m_d\right)}$ can be similarly decomposed using the TT format, in which case it is elementwise written as:
\begin{align}
\tens{A}_{\bm{i}, \bm{j}} = \sum_{r_1=1}^{R_1} \sum_{r_2=1}^{R_2} \cdots \sum_{r_{d-1}=1}^{R_{d-1}} \tens{g}^{(1)}_{1 i_1 j_1 r_1} \tens{g}^{(2)}_{r_1 i_2 j_2 r_2} \cdots \tens{g}^{(d)}_{r_{d-1} i_d j_d 1},
\label{eq:tt_matrix}
\end{align}
where the TT-cores $\tens{g}^{(k)} \in \mathbb{R} ^ {R_{k-1} \times n_k \times m_k \times R_k}$ are now four-dimensional. 
The product between a tensor-operator and a tensor, e.g. the Kronecker product defined in \eqref{eq:tensor-matrix-vector-product}, can be efficiently computed directly in the TT-format \cite{oseledets1}.
\subsection{Linear system solutions in the TT format}
\label{subsec:tt-linear-systems}

Of particular interest in the context of this work is to solve efficiently a multilinear system $\tens{A} \tens{x} = \tens{b}$, where $\tens{A} \in \mathbb{R}^{\left(n_1 \times \cdots \times n_d\right) \times \left(n_1 \times \cdots \times n_d\right)}$ and $\tens{x}, \tens{b} \in \mathbb{R}^{n_1 \times \dots \times n_d}$ are given in the TT format.  
Iterative Krylov subspace solvers, e.g. based on the conjugate gradient (CG) or generalized minimal residual (GMRES) methods, can be generalized to multilinear systems given in the TT format \cite{dolgov2013tt}. 
However, without preconditioning, the number of iterations is large and leads to a large number of rounding operations in order to keep
the TT-rank small, thus resulting in an undesirable computational cost \cite{dolgov_oseledets_2012,dolgov2013tt}. 

An alternative approach is to minimize the norm of the system's residual with respect to the cores of the solution's TT-decomposition.
The corresponding minimization problem reads 
\begin{align}
\label{eq:min_problem}
\min\limits_{\tens{x}\in \mathbb{R}^{n_1 \times \cdots \times n_d}} \left\lVert \tens{A} \tens{x} - \tens{b} \right\rVert_{\rm F}^2,
\end{align}
where $\lVert \cdot \rVert_{\rm F}$ denotes the Frobenius norm.
The feasible set $\mathbb{R}^{n_1 \times \cdots \times n_d}$ is restricted to a subset of $\mathbb{R}^{n_1 \times \cdots \times n_d}$ which contains all $d$-dimensional tensors that can be represented in the TT format with TT-ranks $R_k \leq R_{\max}$, $k=1,\dots,d-1$. 
In this case, the minimization problem is nonlinear with respect to the core tensors $\tens{g}^{(k)}$ of the TT representation of $\tens{x}$ given in \eqref{eq:tt-format}.

The nonlinear minimization problem \eqref{eq:min_problem} can be solved using the alternating least squares (ALS) method \cite{holtz2012alternating, grasedyck2015variants}.  
The method fixes all cores but one, thus resulting in a quadratic optimization problem for the minimizing core. 
The process is then repeated iteratively, minimizing one core at a time until the objective function decreases to a sufficiently small value \cite{dolgov_oseledets_2012}. 
The main drawback of the ALS method is that the TT ranks must be given a priori.
This can be avoided by minimizing over two consecutive cores instead of just one, an idea that was first introduced by the Density Matrix Renormalization Group (DMRG) algorithm \cite{white1993density} and which was later used in the Alternating Minimal Energy (AMEn) method for solving high-dimensional multilinear systems \cite{amen1, dolgov_tamen}.
In order to bring the minimization problem to a quadratic form, a so-called supercore $\widetilde{\tens{g}}^{(k,k+1)} \in \mathbb{R}^{ R_{k-1} \times n_k \times n_{k+1} \times R_{k+1}}$ is first defined as
\begin{equation}
\widetilde{\tens{g}}^{(k,k+1)} = \tens{g}^{(k)} \tens{g}^{(k+1)}, 
\end{equation} 
thus removing all information regarding the rank $R_k$.
Then, the TT-representation of $\tens{x}$ takes the following form (in elementwise notation):
\begin{align}
\tens{x}_{\bm{i}} = \sum\limits_{r_i, i \neq k} \tens{g}^{(1)}_{1 i_1 r_1}\cdots\widetilde{\tens{g}}^{(k,k+1)}_{r_{k-1} i_k i_{k+1} r_{k+1}} \cdots \tens{g}^{(d)}_{r_{d-1} i_d 1}.
\end{align}
The minimization then proceeds similar to the ALS method, where now one supercore is minimized in each iteration.
After the optimization procedure is completed, the supercore is divided into two separate TT-cores, e.g. by means of SVDs of auxiliary matrices \cite{savostyanov2011fast}.
Then, the rank $R_k$ is not a priori given, but identified as part of the supercore's separation.

\subsection{Low-rank TT representation of the CME operator}
The CME operator in \eqref{eq:cme_op} can be represented in the TT format using a sum of rank-$\bm{1}$ tensors, without ever assembling the full tensor-operator \cite{cme_time, hegland2011}. 
For a reaction of type \eqref{eq:reaction} with the propensity function \eqref{eq:props}, one can use the separation 
\begin{subequations}
	\begin{align}
	\alpha_m(\bm{x}) &= c_m f_1^{(m)}(x_1) \cdots f_d^{(m)}(x_d), \\
	f_k^{(m)}\left(x_k\right) &= \frac{x_k!}{q_{m,k}!\left(x_k - q_{m,k}\right)!}. \label{eq:split_prop}
	\end{align}
\end{subequations}
Then, the CME tensor-operator defined in \eqref{eq:cme_op} can be written as the difference between two tensor-matrices $\tens{B}, \tens{C} \in \mathbb{R}^{\left(n_1 \times \cdots \times n_d\right) \times \left(n_1 \times \cdots \times n_d\right)}$, i.e. 
\begin{align}
\tens{A}_{\bm{i}, \bm{j}} = \tens{B}_{\bm{i}, \bm{j}} - \tens{C}_{\bm{i}, \bm{j}},
\end{align}
both of which admit exact rank-$\bm{1}$ TT decompositions, with the corresponding four-dimensional TT-cores $\tens{g}^{(k)}$ (for $\tens{B}$) and $\tens{h}^{(k)}$ (for $\tens{C}$) given by
\begin{subequations}
	\begin{align}
	\tens{g}^{(k)}_{1 i_k j_k 1} &= f_k^{(m)}(x_k^{(i_k)})  \delta_{i_k}^{j_k}  \mathbb{I}_{n_k} (x_k{(j_k)} + \nu_k^{(m)} ), \\
	\tens{h}^{(k)}_{1 i_k j_k 1} &=  f_k^{(m)}(x_k^{(j_k)}) \delta_{i_k - \nu_k}^{j_k} \mathbb{I}_{n_k}(x_k{(j_k)} + \nu_k^{(m)} ),
	\end{align}
\end{subequations}
where the indicator functions $\mathbb{I}_{n_k}$ are defined as $\mathbb{I}_{n_k}(x) = 1$ if $x\leq n_k$ and $0$ otherwise.
Then, for the $m$th reaction, the maximum TT-rank of the tensor-operator $\tens{A}$ is at most equal to 2. 
If the operators coming from $M$ different reactions are added, the maximum TT-rank is at most $2M$. 
However, in practical examples, rank rounding with a very high accuracy, e.g. $\lVert \widetilde{\tens{A}} - \tens{A}\rVert_{\text{F}} \approx 10^{-12}||\tens{A}||_{\text F}$, yields a TT approximation $\widetilde{\tens{A}} \approx \tens{A}$ with much lower TT-ranks.

\subsection{Solving the CME in the TT format}
\label{sec:tamen}
The CME can be solved numerically in the TT-format using finite differences \cite{gelss2017tensor}. 
This limits the choice of the time discretization to classical implicit and explicit schemes\cite{dolgov_tamen}. 
An alternative method \cite{cme_time,dolgov_tamen} is to employ a basis representation of the time-dependent solution over an interval $[0,\Delta T]$, such that
\begin{align}
p(t, \bm{x}(\bm{i})) = \sum\limits_{j=1}^{T} \tens{p}_{\bm{i}j} b_j(t), \label{eq:basis_time}
\end{align}
where $b_j(t)$ are basis functions for the interpolation in time. In this work, we employ a Chebyshev basis, however, other options are also possible, e.g. hat functions or Lagrange polynomials. 

By including time as an additional dimension next to the states, a $(d+1)$-dimensional tensor is obtained.
We then perform a Galerkin projection to recover the degrees of freedom for the entire subinterval, by solving the system
\begin{subequations}
	\begin{align}
	\tens{M} \tens{p} &= \tens{f}, \label{eq:cme-tt-system}\\
	\tens{M}  &=  \tens{I}_{\bm{n}} \otimes (\bm{S}+\bm{V}) - (\tens{I}_{\bm{n}} \otimes \bm{P})(\tens{A} \otimes \bm{I}_T), \label{eq:cme-tt-M} \\
	\tens{f} &=  \tens{p}^{(0)} \otimes \bm{v}, \label{eq:cme-rhs}
	\end{align} 
\end{subequations}
where $\bm{S}$ is the stiffness matrix, $\bm{P}$ is the mass matrix \cite{trefethen2000spectral,dolgov_tamen}, $\bm{I}_N \in \mathbb{R}^{N \times N}$ denotes an identity matrix, and $\tens{I}_{\bm{n}} = \bm{I}_{n_1} \otimes \cdots \otimes \bm{I}_{n_d}$.
The system \eqref{eq:cme-tt-system} can then be solved as described in section \ref{subsec:tt-linear-systems}. 

Due to the increased complexity of the solution over long simulation times, one can divide the time domain and apply the presented method to 
the individual subdomains by taking the end state as initial condition for the next subinterval. 
For a constant subinterval length and if the solution is smooth, the convergence is exponential in $T$\cite{dolgov_tamen,trefethen2000spectral}. An error indicator is represented by computing the norm of the residual of \eqref{eq:cme-tt-system} for an enriched basis\cite{dolgov_tamen}
\begin{align}
\varepsilon(T,\Delta t) = || \tens{M}' \tens{Q} \tens{p} - \tens{f}'||_F,
\end{align}
where $\tens{M}'$ and $\tens{f}'$ are the tensors from \eqref{eq:cme-tt-M} and \eqref{eq:cme-rhs} constructed for $T'=2T$. The operator $\tens{Q}$ interpolates the solution on the finer time-domain basis with $T'=2T$. 
This can be used to adapt the subinterval length if the indicator $\varepsilon(T,\Delta t)$ is larger than a prescribe tolerance tol\cite{dolgov_tamen}
\begin{align}
\Delta t' = \left(\frac{\text{tol}}{\varepsilon(T,\Delta t)}\right)^{\frac{1}{T}} \Delta t,
\end{align}
where $\Delta t'$ is the modified subinterval length.
One bottleneck is the prescribed accuracy of the TT-solver, which, as shown in the numerical results, acts like a lower limit for the error.

In the same framework, one can also include classical implicit time stepping schemes, such as the Crank-Nicolson and the implicit Euler, by appropriately choosing the stiffness and mass matrices \cite{dolgov_tamen}. The motivation behind this is that the time dynamics is captured in the low rank structure of the tensor, reducing the storage complexity. Moreover, as observed from numerical experiments, the runtime time is also reduced.

\subsection{Quantized TT CME solver}

One way to optimize the aforementioned TT-based CME solver is to employ the so-called quantized tensor train (QTT) decomposition\cite{khoromskij2011d}. The QTT format has proven to further increase the storage and computational efficiency of the TT representation by reshaping the tensors into higher dimensional ones while reducing the mode sizes. Let $\tens{x}\in\mathbb{R}^{n_1\times n_2 \times\cdots\times n_d}$ be a tensor with $\log_2 n_k \in\mathbb{N},k=1,...,d$, i.e. with mode sizes that are powers of 2. If the tensor $\tens{x}$ is represented in the TT format, then reshaping it into a $(\sum_k \log_2 n_k)$-dimensional tensor can be easily achieved by performing the TT-decomposition on the individually reshaped cores.

Let $\tens{x} \in \mathbb{R}^{n_1 \times \cdots \times n_d}$ be a tensor with $\log_2 n_k \in \mathbb{N}$. The tensor admits a rank $\bm{R}$ TT-decomposition with the cores $\tens{g}^{(k)}$. The quantization process implies reshaping the individual cores to tensors of shape $r_{k-1}\times2 \times \cdots \times 2 \times r_k $ and then the TT-decomposition of the reshaped cores is computed. The resulting cores correspond to the QTT decomposition of the tensor \cite{oseledets2010tt}. In the case of the tensor-matrices, the procedure is similar. Compared to the computational complexity of the solver, the complexity of the transformation between TT and QTT can be neglected.    
This procedure has been proven to be effective for reducing the storage requirements and the computation time for solving the CME in the TT format\cite{kazeev_2014, cme_time}. 
If the ranks of the QTT decomposition remain bounded, the storage complexity $\mathcal{O}(d\log_2 N)$\cite{qtt_FP}. Moreover, the solver benefits from the linearity with respect to the number of dimensions.

\section{Bayesian Inference for the Chemical Master Equation with Parameter Dependencies}
\label{sec:infer-tt-cme}

\subsection{Parameter-dependent CME}
\label{subseq:param_cme}

We now consider a parameter-dependent CME \cite{cme_time}, described by 
\begin{align}
\frac{\text{d} \tens{p}(\bm{\theta})}{\text{d}t} = \tens{A}\left(\bm{\theta}\right) \tens{p}(\bm{\theta}),
\end{align}
where $\bm{\theta} \in \mathbb{R}^{n_p}$ denotes the parameter vector. 
The parameter vector $\bm{\theta}$ is assumed to take values in the tensor-product space $\mathcal{P} = [\theta_1^{\min}, \theta_1^{\max}] \times \cdots \times [\theta_{n_p}^{\min}, \theta_{n_p}^{\max}]$.

Solving the CME for one parameter realization $\bm{\theta}^{(\bm{l})} \in \Theta$, $\bm{l} = (l_1, \dots, l_{n_p})$, yields the conditional PMF $p_{t}\left(\bm{x}|\bm{\theta}^{(\bm{l})}\right)$. 
If instead of a fixed initial PMF, one starts with the joint PMF $p_{t_0}\left(\bm{x},\bm{\theta}\right)$, the solution will be the joint PMF over the discretized time interval. Since our goal is to compute the conditional/joint PMF over the entire parameter space, we use a basis expansion \cite{bigoni_spectral} to describe the parameter dependence, such that

\begin{align}
p_t\left(\bm{x}(\bm{i}),\bm{\theta}\right) \approx \sum\limits_{j} \sum\limits_{\bm{l}} \tens{p}_{\bm{i}\bm{l}j}(t) L_{\bm{l}}(\bm{\theta}), \label{eq:pmf_param}
\end{align}
where $\tens{p} \in \mathbb{R}^{n_1 \times \cdots \times n_d \times \ell_1 \times \cdots \times \ell_{n_p} }$,  $\{L_{\bm{l}}\}_{\bm{l}=\bm{1}}^{\bm{\ell}}$ is a tensor-product basis  $L_{\bm{l}}(\bm{\theta})=L^{(1)}(\theta_1)\cdots L^{(n_p)}(\theta_{n_p})$.

In this work, B-spline basis functions are chosen for the parameter dependence \cite{boorBsplines}.

In  order to retrieve the degrees of freedom, a Galerkin formulation is used to derive a multilinear system with respect to the full tensor. We choose the test functions $\tens{q}(\bm{\theta})$ from the same space, which yield the formulation
\begin{align}
\langle \frac{\text{d} \tens{p}(\bm{\theta})}{\text{d}t},\tens{q} \rangle = \langle \tens{A} \tens{p} , \tens{q} \rangle,
\end{align}
where $\langle \cdot,\cdot \rangle$ is an inner product with respect to $\bm{\theta}$. One can then derive the multilinear system
\begin{align}
\tens{M} \frac{\text{d} \tens{p}(\bm{\theta})}{\text{d}t} = \tens{K} \tens{p},
\end{align}
with the mass tensor-matrix
\begin{align}
\tens{M}_{\bm{mn},\bm{il}} = \delta_{\bm{m}}^{\bm{i}} \int\limits_\mathcal{P} L_{\bm{n}}(\bm{\theta}) L_{\bm{l}}(\bm{\theta}) \text{d}\bm{\theta},
\end{align}
and the stiffness tensor-matrix
\begin{align}
\tens{K}_{\bm{mn},\bm{il}} =  \int\limits_\mathcal{P} \tens{A}_{\bm{m},\bm{i}} (\bm{\theta})L_{\bm{n}}(\bm{\theta}) L_{\bm{l}}(\bm{\theta}) \text{d}\bm{\theta}.
\end{align}
The mass matrix can be easily constructed as a rank-$\bm{1}$ TT-operator using the individual mass matrices of the univariate bases. On the contrary, the stiffness matrix requires the evaluation of the parameter dependent CME operator over a tensor product quadrature grid $\Theta = \big\{  \theta_1^{\left(r_1\right)} \big\}_{r_1=1}^{\ell_1} \times \big\{  \theta_2^{\left(r_2\right)} \big\}_{r_2=1}^{\ell_2} \times\cdots \times \big\{  \theta_{n_p}^{(r_{n_p})} \big\}_{r_{n_p}=1}^{\ell_{n_p}}$, such that
\begin{align}
\tens{K}_{\bm{mn},\bm{il}} \approx \sum_{\bm{r}} \tens{w}_{\bm{r}} \bar{\tens{A}}_{\bm{m}\bm{r},\bm{i}\bm{r}} L_{\bm{n}}(\bm{\theta}^{(\bm{r})}) L_{\bm{l}}(\bm{\theta}^{(\bm{r})}),
\end{align}
where $\tens{w}$ is the weight tensor and $\bar{\tens{A}}_{\bm{ik},\bm{jr}} = \tens{A}\left(\bm{\theta}^{(\bm{r})}\right) \delta_{\bm{r}}^{\bm{k}}$. The tensor-matrix $\tens{K}$ admits a TT representation or approximation, assuming that the evaluation of the CME operator can also be represented or approximated in the TT format.


A direct construction of the extended operator $\bar{\tens{A}}$ can be easily accomplished if each parameter affects only one reaction. 
For example, this situation occurs if the parameters are the reaction rates, i.e. $\bm{\theta} = (c_1,c_2,\dots)$. 
Then, the operator is extended using the Kronecker product, such that
\begin{align}
\bar{\tens{A}} = \: &\tens{A}^{(1)} \otimes \left(\text{diag}\left(\theta_1^{(1)},...,\theta_1^{(\ell_1)}\right) \otimes \tens{I}_{\ell_2} \otimes \tens{I}_{\ell_3} \otimes\cdots \right) +  \nonumber \\
&\tens{A}^{(2)} \otimes \left( \tens{I}_{\ell_1} \otimes \text{diag}\left(\theta_2^{(1)},...,\theta_2^{(\ell_2)}\right) \otimes \tens{I}_{\ell_3} \otimes \cdots \right) + ... , \label{eq:params_operator}
\end{align}
where $\tens{A}^{(m)}$ is the CME operator corresponding to 
reaction $m$ for a unity reaction rate.

We note that parameter dependence is not necessarily restricted to the reaction rates.  Equation \eqref{eq:params_operator} can be extended to accommodate other types of parameter dependencies. If the propensity functions have a representation as in \eqref{eq:split_prop} and every $\alpha_m$ depends on at most one parameter, then the individual CME operator for every reaction evaluated on the grid $\Theta$ can be expressed as a sum of rank-$\bm{1}$ TT tensors and then rounded to eliminate overshooting ranks. Moreover, even the structure of the reaction network can be incorporated as a parameter, however, this is out of scope for the present work.


\subsection{Filtering and smoothing in the TT format}
\label{subsec:filter}

One relevant inference task in computational biology applications is the filtering and smoothing of observations, for example, in order to estimate the dynamics of genes that are measured indirectly via a fluorescent reporter protein. 
We consider $N_{\rm o}$ state observations $\left\{\bm{y}^{(j)}\right\}_{j=1}^{N_{\text{o}}}$, which are sampled at discrete time steps $\left\{t_j\right\}_{j=1}^{N_{\text{o}}}$. 
The observations are considered to be realizations of the random variables $\left\{\bm{Y}^{(j)}\right\}_{j=1}^{N_{\text{o}}}$, which are assumed to be conditionally independent given the latent states $\left\{\bm{X}(t_j)\right\}_{j=1}^{N_{\text{o}}}$. 
Thus, the observation model is assumed to be dependent only on the current state. Its probability density function (PDF) is denoted with $p_{\bm{Y}|\bm{X}}\left(\bm{y}|\bm{x}\right)$.  In practice, $p_{\bm{Y}|\bm{X}}$ often corresponds to additive Gaussian or multiplicative lognormal noise. However, the presented framework is not limited to these particular cases. 

The conditional probability $\text{Pr}(\bm{X}(t)=\bm{x}|\bm{y}^{(1)},...,\bm{y}^{(j)})$ for $j=\max\{k\in\mathbb{N} | t_k<t\}$ satisfies the unconditional master equation \cite{huang2016reconstructing}
\begin{subequations}
	\begin{align} \label{eq:fwd_cme}
	\frac{{\rm d} p_t(\bm{x})}{{\rm d} t} = \sum_{m=1}^M \left\{ \alpha_m({\bm{x}-\bm{\nu}^{(m)}}) p_t({\bm{x}-\bm{\nu}^{(m)}}) - \alpha_m\left(\bm{x}\right) p_t\left(\bm{x}\right)\right\} , \end{align}
	with the reset conditions
	\begin{align}
	p_{t_j}(\bm{x})=\frac{1}{Z_j}p_{t^{-}_j}(\bm{x})p_{\bm{Y}|\bm{X}}\left(\bm{y}^{(j)}|\bm{x}\right),
	\end{align}
\end{subequations} 
where $p_{t^{-}_j}(\bm{x})$ represents the left approaching limit $t\rightarrow t_j,t<t_j$, and $Z_i=\sum_{\bm{x}}p_{t^{-}_j}(\bm{x})p_{\bm{Y}|\bm{X}}\left(\bm{y}^{(j)}|\bm{x}\right)$. If all observations are taken into consideration, we deal with the smoothing case. The PMF $\tilde{p}_t(\bm{x})$ of $\text{Pr}(\bm{X}(t)=\bm{x}|\bm{y}^{(1)},...,\bm{y}^{(N_{\text{o}})})$ can be factorized as
\begin{align}
\tilde{p}_t(\bm{x}) = p_t(\bm{x}) \beta_t(\bm{x}),
\end{align}
where the PMF satisfies the backward master equation
\begin{subequations}
	\begin{align}
	\label{eq:bck_cme}
	\frac{{\rm d} \beta_t(\bm{x})}{{\rm d} t} &= \sum_{m=1}^M \left\{ \alpha_m({\bm{x}}) \beta_t({\bm{x}}) - \alpha_m(\bm{x}) \beta_t(\bm{x}+\bm{\nu}^{(m)})\right\} , \\ 
	\beta_{t_j^-}(\bm{x})&=\frac{1}{Z_j}p_{t_j}(\bm{x})p_{\bm{Y}|\bm{X}}(\bm{y}^{(j)}|\bm{x}),
	\end{align}\end{subequations}
where $\beta(\bm{x},t_{N_{\text{o}}})=1$ is the terminal condition and  
\begin{align}
\beta_t(\bm{x}) \propto p(\bm{y}^{(j)},...,\bm{y}^{(N_{\text{o}})}|\bm{X}(t)=\bm{x}),
\end{align}
where $j = \min\{k\in\mathbb{N}|t_k>t\}$. The PMF $\tilde{p}_t(\bm{x})$ satisfies the evolution equation 
\begin{align}
\frac{{\rm d} \tilde{p}_t(\bm{x})}{{\rm d} t} = \sum_{m=1}^M \left\{ \tilde{\alpha}_m({\bm{x}-\bm{\nu}^{(m)}},t) \tilde{p}_t({\bm{x}-\bm{\nu}^{(m)}}) - \tilde{\alpha}\left(\bm{x},t\right) \tilde{p}_t\left(\bm{x}\right)\right\}, 
\end{align} 
with the time-varying smoothing propensity functions
\begin{align}
\tilde{\alpha}_m(\bm{x},t) = \alpha_m(\bm{x}) \frac{\beta(\bm{x}+\bm{\nu}^{(m)},t)}{\beta(\bm{x},t)}.
\end{align}

The method is also known in the literature as the forward-backward algorithm \cite{forward_backward} and can be interpreted as a message-passing algorithm in a Hidden Markov Model (HMM). Moreover, it can be efficiently performed in the TT-format, as shown in Algorithm \ref{alg:hmm}.
The PMF $p(\bm{x}^{(j)}|\bm{y}^{(0)},...,\bm{y}^{(j-1)})$ is the forward message and is denoted by the tensor $\tens{a}^{(j)}\in\mathbb{R}^{n_1\times\cdots\times n_d}$. 
The prediction step is performed by solving the CME over the interval $[t_{j-1},t_j]$ with the initial condition $\tens{a}^{(j-1)}$. 
The observation is used to construct a tensor $\tens{p}^{\rm obs}\in \mathbb{R}^{n_1\times\cdots\times n_d}$ with $\tens{p}_{\bm{i}}^{\rm obs} = p_{\bm{Y}|\bm{X}}\left(\bm{y}^{(j)}|\bm{x}(\bm{i})\right)$. 
If every species is observed independently, i.e. the observation model can be factorized, the tensor $\tens{p}^{\rm obs}$ is rank-$\bm{1}$ and can be expressed as a Kronecker product. 
For the backward pass, the message is $p\left(\bm{y}^{(j+1)},...,\bm{y}^{(N_{\rm o})}|\bm{x}^{(j)}\right)$ and is represented with the tensor $\tens{b}^{(j)}$. 
The CME is now solved using the transposed operator $\tens{A}^\top$ and with the initial condition $\tens{b}^{(j+1)} \ast \tens{p}^{\rm obs}$, where $\ast$ denotes the elementwise multiplication operation. 
The last step is to multiply and normalize the forward and the backward messages in order to get the conditional $p\left(\bm{x}^{(j)}|\bm{y}^{(0)},...,\bm{y}^{(N_{\rm o})}\right)$. 
In the presented framework, we a get smoother distribution only at the observation points. In order to have information in between the observations, prediction steps must be added.

\begin{algorithm}[H]
	\begin{algorithmic}
		\STATE \textbf{Input:} Sample $\{\bm{y}^{(j)}\}_{j=0}^{N_{\text{o}}}$, initial PMF $\tens{p}^{(0)}$
		\STATE $\tens{a}^{(0)} \gets \tens{p}^{(0)}$
		\FOR {$j=1,...,N_{\text{o}}$} 
		\STATE Solve the CME with $\tens{a}^{(j-1)}$ as initial condition.
		\STATE Compute $\tens{p}^{\rm obs}$ for $\bm{y}^{(j)}$ in the TT format.
		\STATE $\tens{a}^{(j)} \gets\tens{p}^{\rm obs}_{\bm{i}}\ast\tens{a}^{(j-1)} $
		\ENDFOR 
		\STATE $\tens{b}^{(N_{\rm o})} \gets \tens{1} $
		\FOR {$j=N_{\rm{o}} - 1,\dots,0$} 
		\STATE Compute $\tens{p}^{\rm obs}$ for $\bm{y}^{(j+1)}$ in the TT format.
		\STATE Solve the CME with operator $\tens{A}^\top$ and initial condition $Z^{-1}\tens{b}^{(j+1)}\ast\tens{p}^{\rm obs}$.
		\STATE $\tens{b}^{(j)} \gets\tens{p}^{\rm obs}\ast\tens{b}^{(j+1)} $
		\ENDFOR 
		\FOR {$j=0,...,N_{\rm o}$} 
		\STATE $\tens{p}^{(j)} \gets Z^{-1} \tens{a}^{(j)}\ast\tens{b}^{(j)}$ 
		\ENDFOR
		\STATE \textbf{Output: } $\tens{p}^{(j)}$ for $j=0,...,N_{\rm o}$
	\end{algorithmic}
	\caption{Forward-backward algorithm in the TT format.}
	\label{alg:hmm}
\end{algorithm}

\subsection{Bayesian parameter inference in the TT format}
\label{subsec:param-id}


Consider an observation sample $\left\{\bm{y}^{(j)}\right\}_{j=1}^{N_{\rm o}}$ satisfying the assumptions detailed in section \ref{subsec:filter}, but now connected to a realization of the random process $\bm{X}(t,\hat{\bm{\theta}})$, where $\hat{\bm{\theta}}$ is the parameter vector governing the system. Given a prior distribution $p(\bm{\theta})$, we are interested in computing the Bayesian parameter posterior $ p( \bm{\theta} | \bm{y}^{(0)},...,\bm{y}^{(N_{\rm o})})$.
By viewing the parameters as part of an augmented process $\{\mathbf X(t), \bm{\theta}(t)\}_{t \geq 0}$, the distribution of the parameters over the parameter space $\mathcal{P}$ can be obtained by performing filtering over the joint space of states and parameters. 
The prediction step is given by
\begin{eqnarray}
p\left(\bm{x}^{(j)},\bm{\theta}^{(j)}|\bm{y}^{(0)},...,\bm{y}^{(j-1)}\right) = \nonumber \\ \sum\limits_{x^{(j-1)}} \int \Big\{ p_{j|j-1}\left(\bm{x}^{(k)},\bm{\theta}^{(j)}|\bm{x}^{(j-1)},\bm{\theta}^{(j-1)}\right) \nonumber \\
p\left(\bm{x}^{(j-1)},\bm{\theta}^{(j-1)}|\bm{y}^{(0)},...,\bm{y}^{(j-1)}\right)\Big\} \text{d}\bm{\theta}^{(j-1)},
\end{eqnarray}
where $p_{j|j-1}$ is the transition PDF and implies solving the parameter-dependent CME from $t_{j-1}$ to $t_j$. 
Next, the update step reads
\begin{eqnarray}
p\left(\bm{x}^{(j)},\bm{\theta}^{(j)}|\bm{y}^{(0)},...,\bm{y}^{(j)}\right) = \frac{1}{Z} p_{\bm{Y}|\bm{X}}\left(\bm{y}^{(j)}|\bm{x}\right) \nonumber \\ p\left(\bm{x}^{(j)},\bm{\theta}^{(j)}|\bm{y}^{(0)},...,\bm{y}^{(j-1)}\right).
\end{eqnarray}

In the TT-format, this parameter inference procedure can be implemented as follows. 
The posterior $p\left(\bm{x}^{(j)},\bm{\theta}^{(j)}|\bm{y}^{(0)},...,\bm{y}^{(j)}\right)$ is represented by the tensor $\tens{p}\in\mathbb{R}^{n_1\times\cdots\times n_d\times\ell_1\times\cdots\times\ell_{n_p}}$, such that 
\begin{align}
p\left(\bm{x}^{(j)},\bm{\theta}^{(j)}|\bm{y}^{(0)},...,\bm{y}^{(j)}\right) =\sum\limits_{\bm{l}} \tens{p}^{(j)}_{\bm{x}\bm{l}} L_{\bm{l}}\left(\bm{\theta}\right).
\end{align}
The prediction step involves solving the parameter-dependent CME with $p\left(\bm{x}^{(j)},\bm{\theta}^{(j)}|\bm{y}^{(0)},...,\bm{y}^{(j)}\right)$ as initial condition, returning the predicted PMF $\tens{p}^{(\rm pred)}$ as a result. 
The resulting tensor is multiplied with the observation tensor at step $j+1$ and normalization is performed to get the new joint distribution
\begin{align}
\tens{p}^{(j+1)}_{\bm{x}\bm{l}} = Z^{-1} \tens{p}^{\rm obs}_{\bm{x}} \tens{p}^{(j \rightarrow j+1)}_{\bm{x}\bm{l}},
\end{align}
where $Z = \sum\limits_{\bm{il}} \tens{p}^{(j+1)}_{\bm{il}} \tens{w}_{\bm{l}}$ is the normalization constant and comes from numerically integrating over the parameter space with the integration weights tensor $\tens{w}$.  
A prior distribution $p\left(\bm{\theta}^{(0)}\right)$ and an exact knowledge of state for $j=0$ is used for the first step, where $p\left(\bm{x}^{(0)},\bm{\theta}^{(0)}|\bm{y}^{(0)}\right) = p\left(\bm{x}^{(0)})p(\bm{\theta}^{(0)}\right)$. 
Once all observations have been used, the state is marginalized to obtain the posterior over the parameter space as
\begin{align}
p\left(\bm{\theta}^{(j)}|\bm{y}^{(0)}...,\bm{y}^{(j)}\right) = \sum\limits_{\bm{x}^{(j)}} p\left(\bm{x}^{(j)},\bm{\theta}^{(j)}|\bm{y}^{(0)},...,\bm{y}^{(j)}\right),
\end{align}
which is computationally efficient if performed in the TT-format.
The procedure is summarized in Algorithm~\ref{alg:param-id}.

\begin{algorithm}[H]
	\begin{algorithmic}
		\STATE \textbf{Input:} Sample $\left\{\bm{y}^{(j)}\right\}_{j=0}^{N_{\rm o}}$, initial PMF $\tens{p}^{(0)}$, prior over the parameter space $\tens{p}^{\rm prior}$
		\STATE $\tens{p}^{(0)} \gets \tens{p}^{(0)} \ast \tens{p}^{\rm prior}$
		\FOR {$j=1,...,N_{\rm o}$} 
		\STATE Solve the CME with $\tens{p}^{(j-1)}$ as initial condition to obtain the solution $\tens{p}^{(j\rightarrow j+1)}$.
		\STATE Compute $\tens{p}^{\rm obs}$ for $\bm{y}^{(j)}$ in TT.
		\STATE $\tens{p}^{(j+1)}_{\bm{i}\bm{l}} \gets  \tens{p}^{\rm obs}_{\bm{i}} \tens{p}^{(j \rightarrow j+1)}_{\bm{i}\bm{l}}$
		\STATE $\tens{p}^{(j+1)} \gets Z^{-1} \tens{p}^{(j+1)}$ for $Z = \sum\limits_{\bm{il}} \tens{p}^{(j+1)}_{\bm{il}} \tens{w}_{\bm{l}}$
		\ENDFOR 
		\STATE $\tens{p}^{\rm post}_{\bm{l}} \gets Z^{-1} \sum\limits_{\bm{i}}  \tens{p}^{(N_{\rm o})}_{\bm{il}} $
		\STATE \textbf{Output: } $\tens{p}^{\rm post}$
	\end{algorithmic}
	\caption{Parameter identification for the parameter-dependent CME in the TT-format.}
	\label{alg:param-id}
\end{algorithm}

\section{Numerical Results}
\label{sec:num-results}

The following numerical experiments aim to showcase the advantages of the proposed framework in terms of accuracy and computational efficiency.
With respect to the latter, storage requirements and computation times are reported for every individual test case. 
All tests were run on a workstation with a 10-core Intel Xeon processor with 64GB of RAM. For the TT operations, the \emph{ttpy} Python package was used in combination with the Intel MKL library. 

\subsection{Validation of the TT-based CME solver}

\subsubsection{Two-dimensional simple gene expression model}
\label{subsubsec:simple-gene}
We first validate the developed TT ODE solver and perform a convergence study based on the two-dimensional simple gene expression model \cite{alberts2002molecular}. 
The four reactions are presented in Table \ref{tab:pp}. 
The initial state is $\bm{x}^{(0)}=(2,4)^\top$ with probability 1. 
The CME is solved in the time interval $[0,1024]$ with a subinterval size of $128$, where arbitrary time units are used. 

\begin{table}[t]
	\caption{Reactions of the simple gene expression model.}
	\label{tab:pp}
	\begin{ruledtabular}
		\begin{tabular}{clll}
			Reaction & $\alpha_m(\bm{x})$ & Rates $c_i$ & Description\\
			\hline
			$\text{mRNA} \rightarrow \emptyset$ & $c_1x_1$  & $0.002$ & mRNA degradation \\
			$\text{mRNA}\rightarrow \text{mRNA}+\text{Protein}$ & $c_2x_1$ & $0.015$ & Translation \\
			$\emptyset\rightarrow \text{mRNA}$ & $c_3$ & $0.1$ & Transcription \\ 
			$\text{Protein}\rightarrow\emptyset$ & $c_4x_2$ & $0.01$ & Protein degradation  \\
		\end{tabular}
	\end{ruledtabular}
\end{table}

The first validation test concerns the maximum relative error of the solution to the CME, computed with the method presented in \ref{sec:tamen}, in dependence to the time dimension $t$ of the basis representation from \eqref{eq:basis_time} inside one subinterval. 
The maximum relative error is given as $\max \left|p_{t_{\rm end}}^{\rm (ref)}(\bm{x}) - p_{t_{\rm end}}(\bm{x})\right| / \max \left|p_{t_{\rm end}}(\bm{x})\right|$, where $t_{\rm end} = 1024$ and the reference solution $p_{t_{\rm end}}^{\rm (ref)}(\bm{x})$ is computed by numerically solving the CME without the TT-decomposition for a very fine time grid.
We note that no truncation of the TT-rank was performed during this validation test, and the relative residual which signifies the convergence of the TT-solver was set to $10^{-13}$. 

\begin{figure}[t!]
	\centering
	\includegraphics{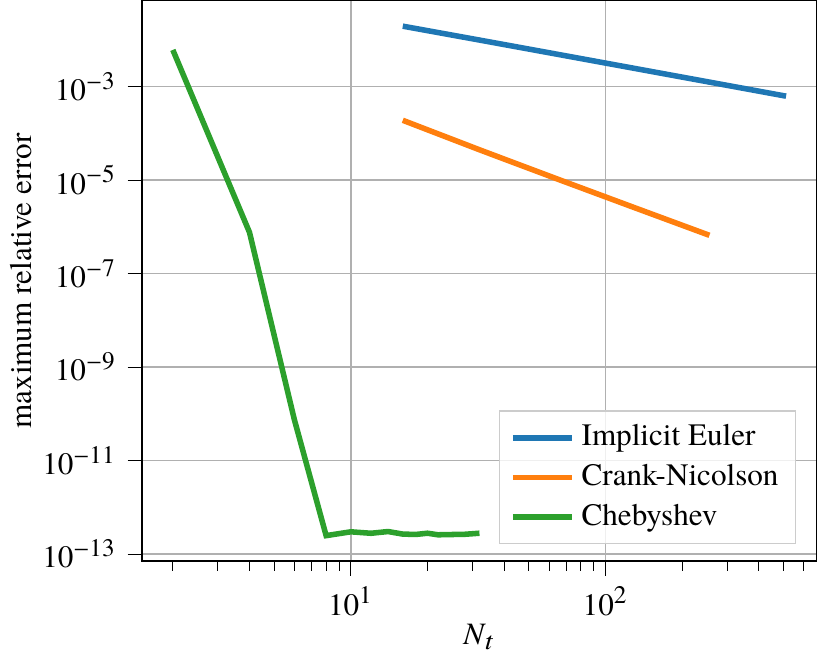}
	\caption{Convergence of the TT-solver with respect to the dimension of the time basis.}
	\label{fig:tt-convergence-euler-cc}
\end{figure}

The results of this first validation set are presented in Fig.~\ref{fig:tt-convergence-euler-cc}, where the employed time-interpolation on the Chebyshev polynomial basis, as shown in \eqref{eq:basis_time}, is compared against classical time-stepping methods such as implicit Euler and Crank-Nicolson finite-difference schemes.  
As would be expected, irrespective of the time-stepping method, the TT-based CME solver yields increasingly more accurate results for finer discretizations of the time interval.
Moreover, as expected from theory, the convergence of the explicit Euler scheme is $\mathcal{O}(\Delta t)$, accordingly $\mathcal{O}(\Delta t^2)$ for Crank-Nicolson \cite{ode_book}.
In the case of the Chebyshev polynomials, an exponential convergence is observed and the error stagnates for a basis of only $T=8$ polynomials.

\begin{figure}[t!]
	\centering
	\includegraphics{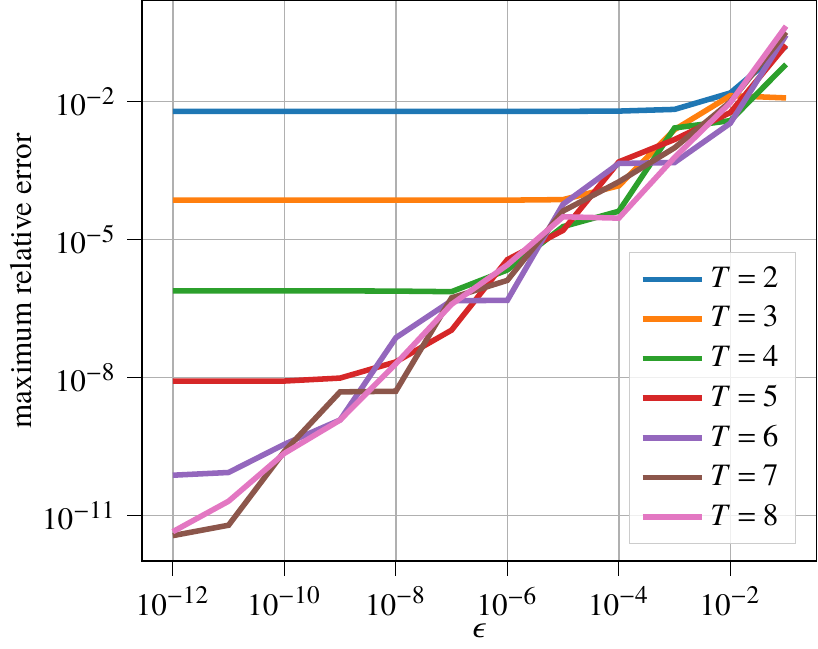}
	\caption{Error versus solver accuracy for different sizes of the basis. }
	\label{fig:tt-convergence-mix}
\end{figure}

The combined impact of the time step and the maximum residual of the TT solver is investigated next for the Chebyshev basis representation, with the results presented in Fig.~\ref{fig:tt-convergence-mix}.  
As can be observed, for a fixed $T$, the accuracy of the TT-solver's solution stagnates after a certain value of the maximum residual.
The stagnation point is actually dependent on the value of $T$, i.e. smaller $T$ values allow for smaller maximum residuals.
Hence, the step size and maximum residual must be chosen according to the desired accuracy of the TT-solution, also taking into consideration the related computational cost.

\subsubsection{Four-dimensional SEIR model}
\label{subsubsec:seir}
For the previously considered two-dimensional model, the reference solution could easily be obtained using an ODE solver. 
We now consider a four-dimensional virus spreading model, namely, the SEIR model \cite{seir_model}, in which case standard ODE solution methods result in high computational demands in terms of computation time and storage needs. 

The individuals of the virus spreading model are separated into four distinct categories: 
1) Susceptible ($S$), i.e. individuals who may become infected; 2) Exposed ($E$), i.e. infected individuals who are not yet contagious; 3) Infected ($I$), i.e. infected individuals who are contagious; 4) Recovered ($R$), i.e. individuals with immunity to the virus. 
The interactions between the individuals are described by the reactions presented in Table \ref{tab:seir}. 
The initial condition is  $\bm{x}(0)=(50,4,0,0)^\top$ and the state space is truncated to $\bm{n}=\left(n_1, n_2, n_3, n_4\right) = (128,128,64,64)$. 
The simulation was performed over the interval $[0,8]$. 
For the TT-solver, the subinterval length is equal to $0.5$ and the subinterval basis dimension is $T=8$.
The reference solution is computed by numerically solving the CME without the TT-decomposition for a very fine time grid.
\begin{table}
	\caption{Reactions of the SEIR model.}
	\label{tab:seir}
	\begin{ruledtabular}
		\begin{tabular}{clll}
			Reaction & $\alpha_m(\bm{x})$ & Rate $c_i$ & Description\\
			\hline
			$S+I \rightarrow E+I$ & $c_1x_1x_3$  & $0.1$ & Susceptible becomes exposed \\ 
			$E \rightarrow I$ & $c_2x_2$  & $0.5$ & Exposed becomes infected \\ 
			$I \rightarrow S$ & $c_3x_3$  & $1.0$ & Infected recovers without immunity \\ 
			$S \rightarrow \emptyset$ & $c_4x_1$  & $0.01$ & Susceptible dies \\ 
			$E \rightarrow \emptyset$ & $c_5x_2$  & $0.01$ & Exposed dies  \\ 
			$I \rightarrow R$ & $c_6x_3$  & $0.01$ & Infected recovers with immunity \\ 
			$\emptyset \rightarrow S$ & $c_7$  & $0.4$ & New susceptible individuals arrive \\ 
		\end{tabular}
	\end{ruledtabular}
\end{table}
\begin{figure*}[t!]\centering
	
	\begin{subfigure}{0.49\linewidth}%
		\centering
		\includegraphics{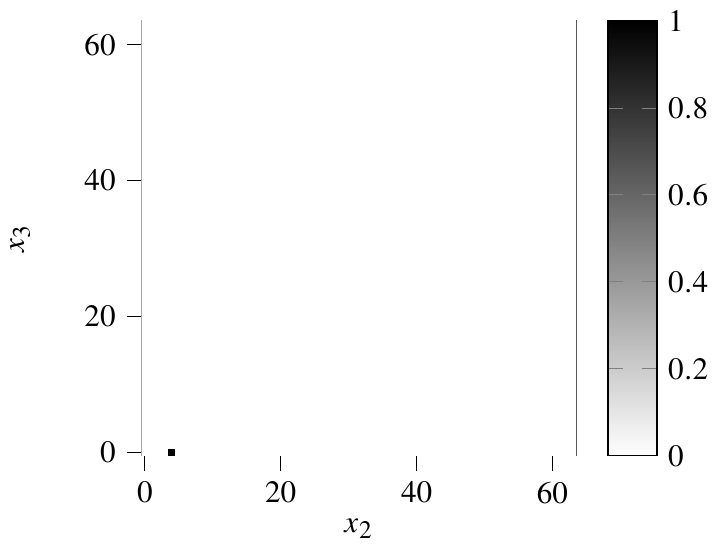}
		\caption{$t=0$.}
	\end{subfigure}
	\begin{subfigure}{0.49\linewidth}
		\centering
		\includegraphics{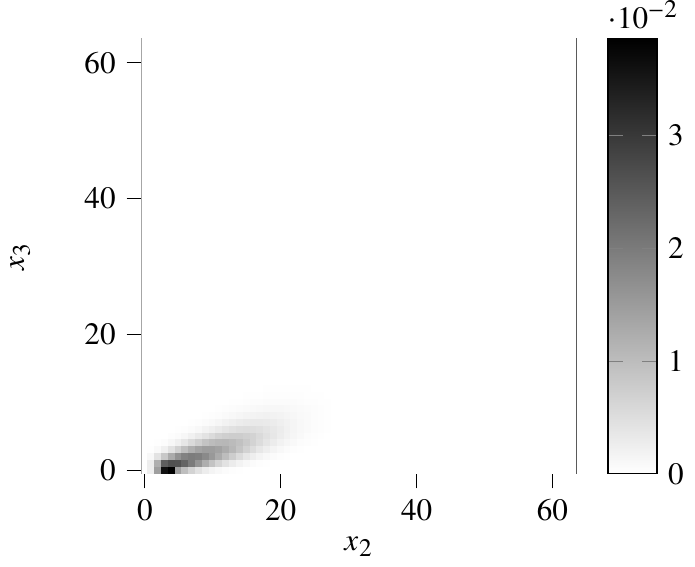}
		\caption{$t=2$.}
	\end{subfigure}
	
	\begin{subfigure}{0.49\linewidth}
		\includegraphics{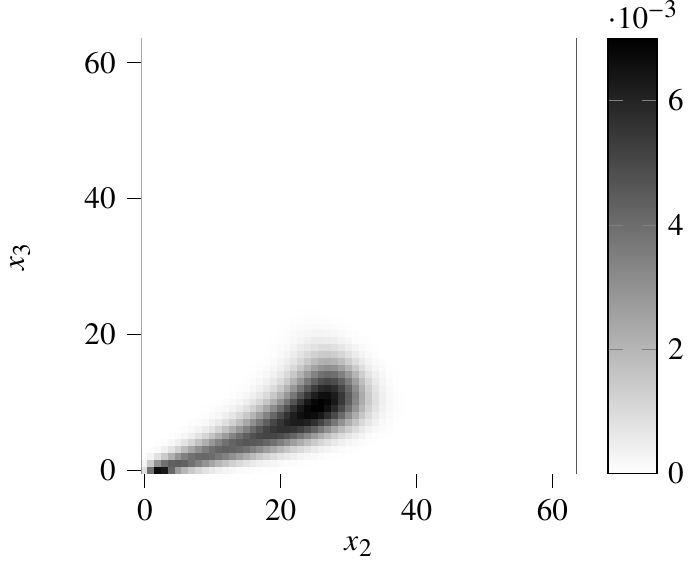}
		\caption{$t=4$.}
	\end{subfigure}
	\begin{subfigure}{0.49\linewidth}
		\includegraphics{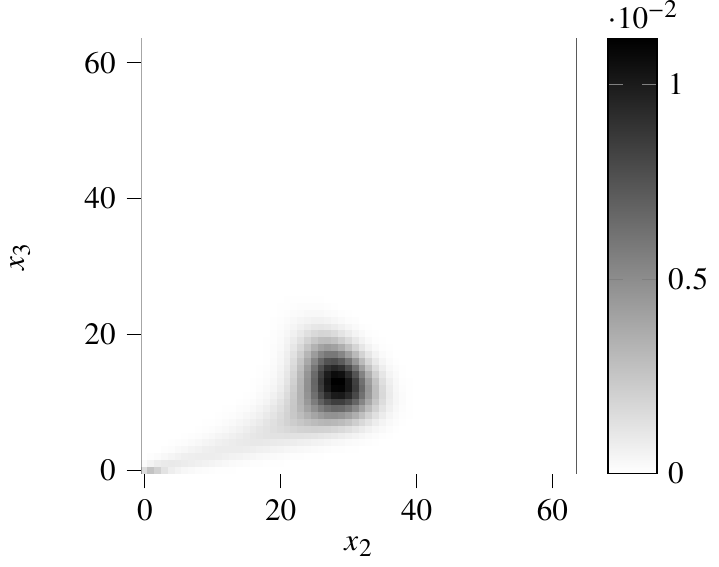}
		\caption{$t=6$.}
	\end{subfigure}
	
	\begin{subfigure}{0.49\linewidth}
		\includegraphics{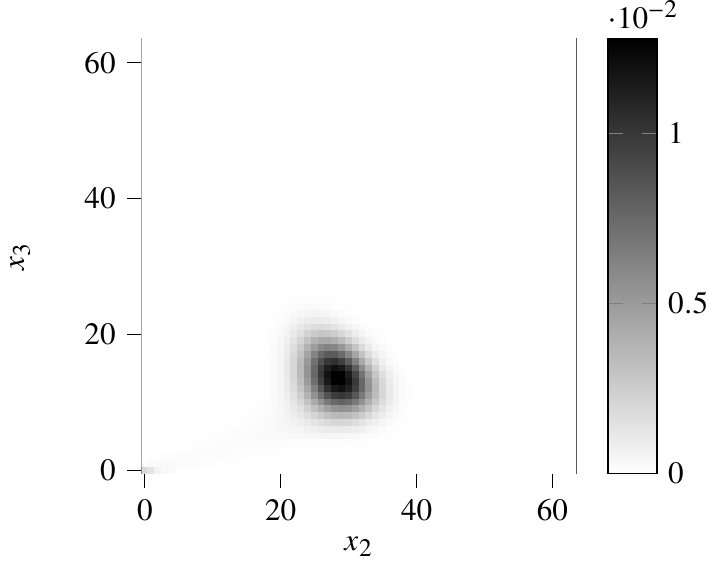}
		\caption{$t=8$.}
	\end{subfigure}
	\begin{subfigure}{0.49\linewidth}
		\includegraphics{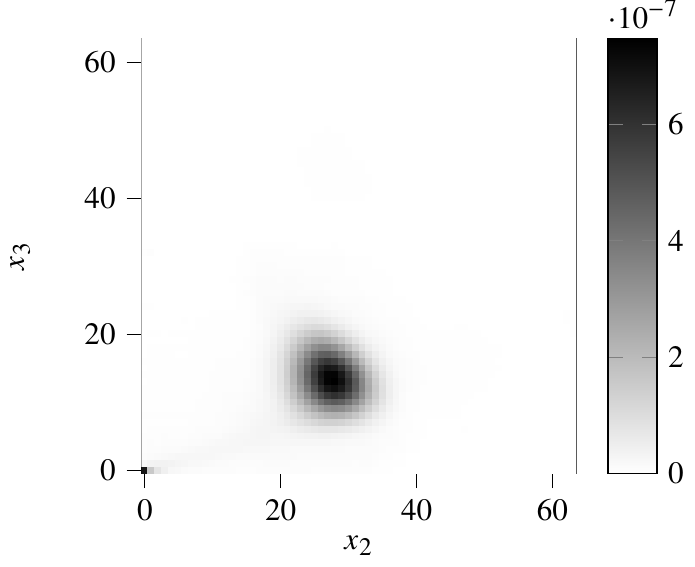}
		\caption{Error at $t=8$.}
	\end{subfigure}
	\caption{Time evolution of Exposed-Infected ($EI$) marginal distribution at $t\in\{0,2,4,6,8\}$ and pointwise absolute error at $t_{\rm end} = 8$. The solution is computed in the TT-format and the reference is obtained by integrating the CME over a fine time grid. }
	\label{fig:ei_plots}
\end{figure*}

Even if a sparse format is employed, $\approx 1.3$~GB RAM are needed to store the CME operator for the reference solution. 
In comparison, using the TT-format, the CME TT-operator has the TT-ranks $\bm{R} = (1,5,6,3,1)$, resulting in storage needs of only $\approx 2.32$~MB RAM, i.e. $0.17 \%$ of the storage space needed by the standard solver. 
If the operator is reshaped in the QTT format, the storage requirements decreases to $\approx 42$ KB.
Moreover, using the solver with the QTT format, the solution is obtained in $\approx 180$~s, which is a considerably smaller computation time than the one needed for the reference solution, i.e. $\approx 12600$~s.
Without the use of the QTT format, the number of solver iterations and the computation time increase by one order of magnitude.

Fig. \ref{fig:ei_plots} shows the time evolution of the marginal $EI$ distribution, as well as the pointwise error at the end of the simulation compared to the reference marginals.
Finally, at $t_{\rm end}$ we obtain the relative errors
\begin{align*}
\epsilon_{\max} = \frac{\max_{\bm{x}} |p^{(\text{ref})}_{t_{\rm end}}(\bm{x})-p_{t_{\rm end}}(\bm{x})|}{\max_{\bm{x}} |p^{(\text{ref})}_{t_{\rm end}}(\bm{x})|} =  2.9 \cdot 10^{-5},
\end{align*}\begin{align*}
\epsilon_{\rm mean} =  \frac{\frac{1}{N^4} \sum\limits_{\bm{x}} |p^{(\text{ref})}_{t_{\rm end}}(\bm{x})-p_{t_{\rm end}}(\bm{x})|}{\max\limits_{\bm{x}} |p^{(\text{ref})}_{t_{\rm end}}(\bm{x})|} = 2.539 \cdot 10^{-9}.
\end{align*}
Hence, the TT-solver yields accurate solutions at significantly reduced execution times and with tremendous storage savings compared to the standard solver.  Indicatively, the solution at $t=8$ requires only $2.5$ MB storage, which is approximately $0.4\%$ the storage requirement of the reference solution.

One issue is the ordering of the species. If species that are highly correlated are apart from each other in the train, the ranks in between must carry the information and therefore the overall rank structure increases. This can also be observed in the representation of the CME operator. In this example, the S,E,I,R ordering is chosen so that most of the reactions involve species that are neighbors in the chain.

\subsection{Filtering and smoothing}
\begin{figure}[ht!] 
	\centering
	\begin{subfigure}{0.99\linewidth}
		\includegraphics{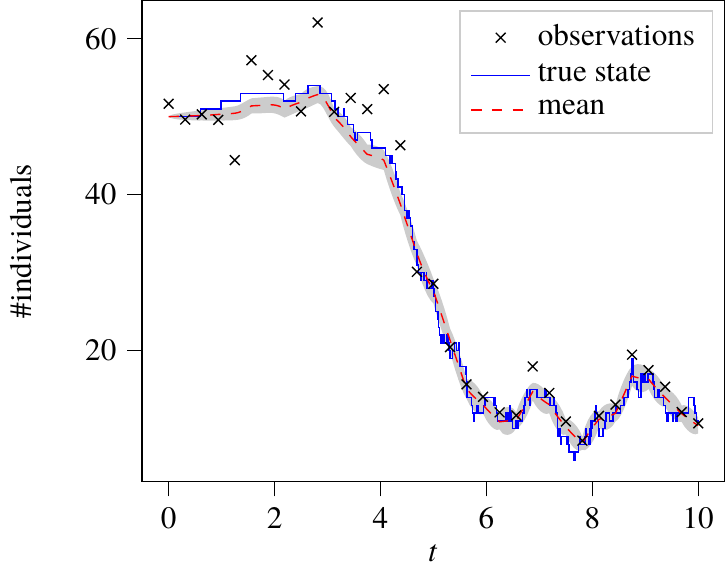}
		\caption{Evolution of the number of susceptible individuals.}
		\label{fig:seir_filter_S}
	\end{subfigure}
	\begin{subfigure}{0.99\linewidth}
		\includegraphics{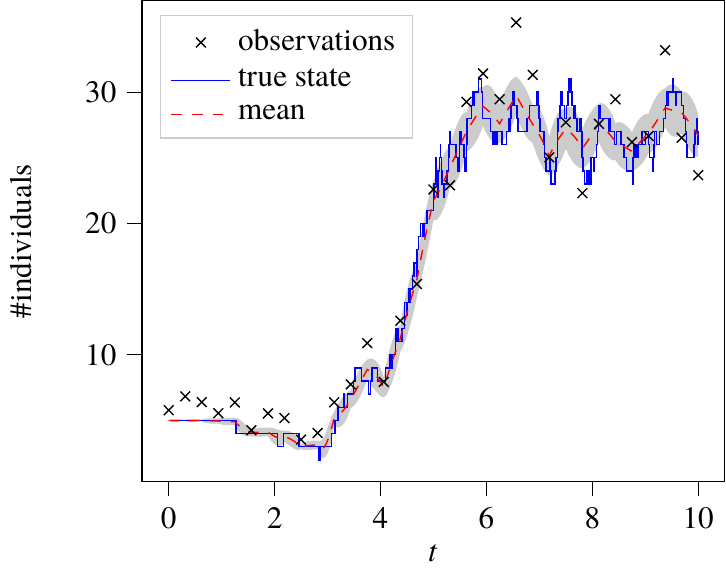}
		\caption{Evolution of the number of exposed individuals.}
		\label{fig:seir_filter_E}
	\end{subfigure}
	\begin{subfigure}{0.99\linewidth}
		\includegraphics{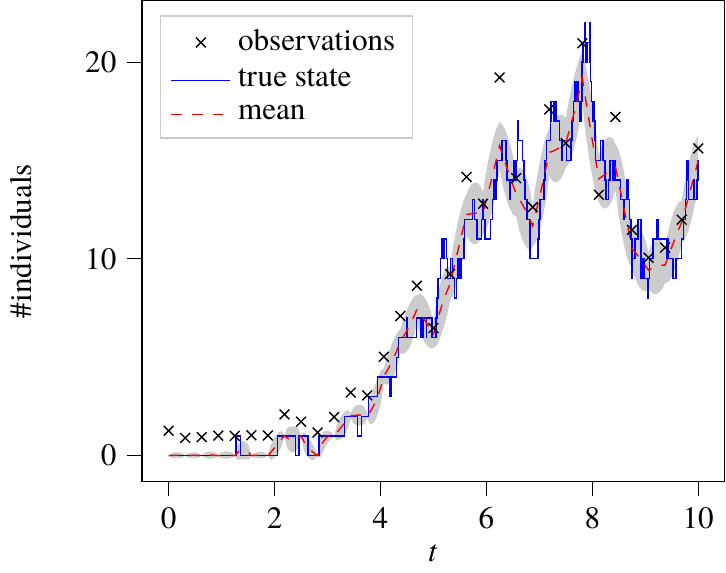}
		\caption{Evolution of the number of infected individuals.}
		\label{fig:seir_filter_I}
	\end{subfigure}
	%
	\caption{Smoothing for the SEIR model with initial population $\bm{x}=(50,2,1,0)^\top$. The sample path is given by the blue line, the observations are marked with ``$\times$'', and for the smoothed distribution the mean (red dashed line) and the standard deviation (gray envelope) are plotted. }
	\label{fig:seir_filter}
\end{figure}

As discussed in \ref{subsec:filter}, state filtering and smoothing can be performed in the TT-framework. 
We consider here the SEIR model presented in section \ref{subsubsec:seir}. 
We assume $N_{\rm o} = 33$ equidistant observations with $\Delta t = 0.3125$. 
The time interval is now chosen as $\left[0,10\right]$.
The realizations are obtained using the Stochastic Simulation Algorithm (SSA) \cite{gillespie_ssa} (blue continuous lines in  Fig. \ref{fig:seir_filter}). 
The noise is assumed to be lognormal with variance $0.1$ for $S$, $E$, and $I$, and $0.05$ for $R$ (black $\times$ symbol in  Fig.~\ref{fig:seir_filter}).

\begin{figure}[ht!] 
	\includegraphics{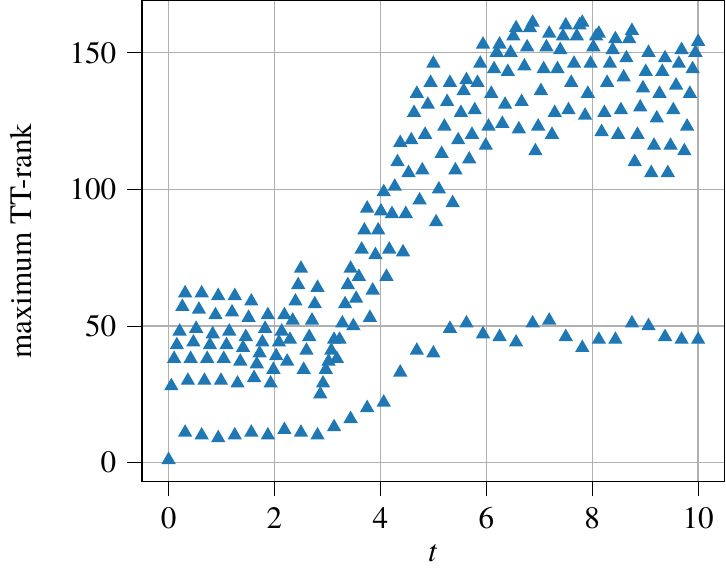}
	\caption{Ranks for the forward pass over simulation time (triangle markers).}
	\label{fig:ranks}
\end{figure}

The TT-based forward-backward Algorithm~\ref{alg:hmm} presented in section \ref{subsec:filter} is used to perform state filtering and smoothing.The state is truncated to $(128,128,64,32)$ and the Chebyshev basis is used for the time dependency.
The runtime of the SEIR experiment is 12 minutes for a solver accuracy of $10^{-6}$ in terms of relative TT-solver residual.
As the estimated state, we compute the expected value of the distribution given by $p\left(\bm{x}^{(k)}|\bm{y}^{(0)},...,\bm{y}^{(N_{\rm o})}\right)$ (red discontinuous line in Fig.~\ref{fig:seir_filter}) and the corresponding standard deviation (grey envelope in Fig.~\ref{fig:seir_filter}). In this case, the incorporation of the observation model in the TT framework is beneficial to the reduction of the error since it acts like a reset condition. 
This can be observed in the decrease of the TT-rank after the multiplication with the tensor corresponding to the observation operator. The numerical experiment shows a decrease in the rank of up to 3 times, as can be observed in Fig. \ref{fig:ranks}.

One more significant advantage of the TT representation is that the storage of the messages decreases dramatically compared to the full tensor representation, as the total storage needed is $\approx 150$ MB for the forward propagating messages and $\approx 190$ MB for the backward propagating messages.
In addition to that, computing the moments of the smooth distribution consists of multiplication with rank-$\bm{1}$ TT tensors.
Moreover, the lognormal observation model is also translated to a rank-$\bm{1}$ tensor. The presented results are performed in the QTT format, however the same test was performed without quantization. For the given state truncation, using the QTT format results in an acceleration by more than an order of magnitude in computation time, compared to the standard TT-format.

\color{black}


\subsection{Parameter inference}

\subsubsection{Simple gene expression}
We now use Algorithm~\ref{alg:param-id} to identify all four parameters $\bm{\theta} = (\theta_1,\theta_2,\theta_3,\theta_4) = (c_1, c_4, c_3, c_4)$ of the simple gene expression model from a noisy sample with $N_{\rm o} = 64$ observations.
In this case, the solver uses the QTT format, the observations are taken equidistantly every $4$ time units, and the parameter priors are independent, truncated Gamma distributions, chosen such that they do not match the actual parameter. 
The parameter domain $\mathbb{R}^4_+$ is restricted to $\theta_i\in[0,6c_i]$. 
As a reference, a sample of size $5\cdot 10^5$ is drawn from the posterior using the Metropolis–Hastings algorithm. 
The CME in this case is solved in the full format with the built-in Python ODE solver. 

With respect to the parameter dependence approximation, the basis of choice in this case is quadratic B-splines with equidistant knots scaled to the parameter range. The dimension of the individual univariate bases is $64$. For the time integrator, a Chebyshev basis of dimension $T=8$ is used with an subinterval size of $0.5$ time units. The runtime is in this case $\approx 21$ minutes with a maximum storage requirement of $\approx 9.2$ MB for the joint over state and parameters (represented by a $6$D tensor with mode size $64$).
The storage requirement for the extended CME operator in the QTT format is only $128$ KB. 
Storing the tensor in the full format is intractable on standard machines even for this $2$D example.
For the given sample size, the Metropolis-Hastings algorithm run for $\approx 1.5$ days.

\begin{figure*}[ht!]
	\centering
	\includegraphics[width=1\textwidth]{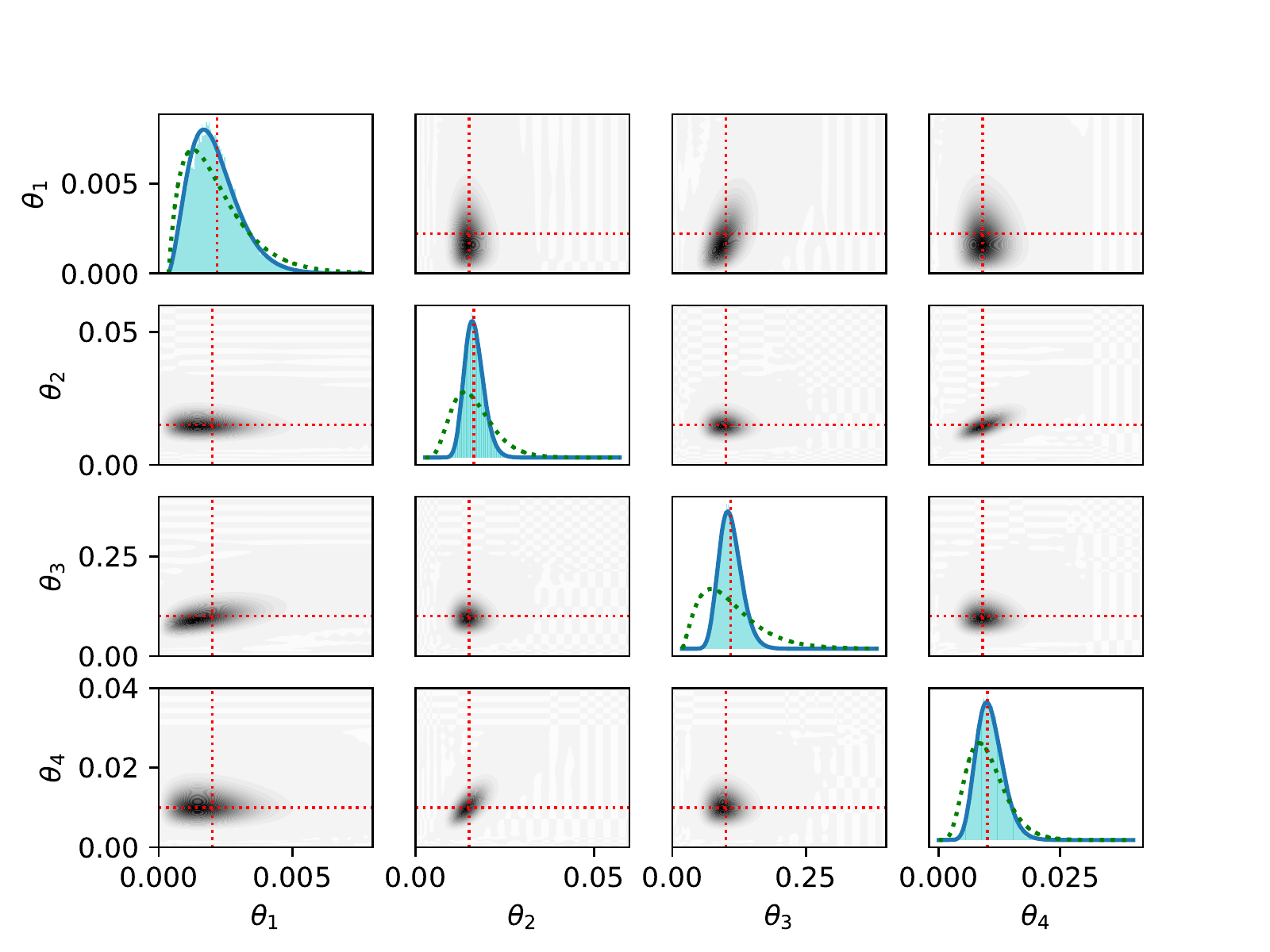}
	\caption{Posterior marginal distributions for the four unknown reaction rates of the simple gene expression model. The black regions correspond to high density of the PDF. The exact parameters are marked with the red dashed lines. For the 1D marginals, a histogram of the posterior sample is represented as a reference, as well as the prior (green dashed lines). }
	\label{fig:filter_2}
\end{figure*}

Since the posterior over the parameter space is four-dimensional, hence, not easy to visualize, the marginals for the individual parameters are computed and compared to the $1$D histograms of the posterior sample, for the purpose of validation. 
The results are presented in Fig.~\ref{fig:filter_2}, where it can be observed that the posterior modes offer a reasonable approximation of the true values, the latter being marked with red vertical dashed lines.
As a further characterization of the posterior, we compute its expected value, variance, and the mode of the PDF:
\begin{align*}
\mathbb{E}[\bm{\theta}] &= (0.001924, 0.01512, 0.09985, 0.01057),\\
\mathbb{V}[\bm{\theta}] &= (1.034\times10^{-6}, 8.685\times10^{-6}, 5.428\times10^{-4}, 7.624\times10^{-6}), \\
\hat{\bm{\theta}} &= (0.001373, 0.01412, 0.09065, 0.009567).
\end{align*}
For comparison, the mean and variance of the reference posterior sample are:
\begin{align*}
\bm{\mu}_{\bm{\theta}} &= (0.001922, 0.01507, 0.09992, 0.01052),\\
\bm{\sigma}_{\bm{\theta}}^2 &= (9.975\times10^{-7}, 8.498\times10^{-6}, 5.233\times10^{-4}, 7.518\times10^{-6}).
\end{align*}

Since there is no analytical estimate for the combined error of the method, several runs with different hyperparameters are performed for this model. First, the accuracy of the solver in terms of relative residual, here denoted with $\epsilon$, is varied and the relative error of the TT-solver's solution is analyzed, first with respect to the MCMC solution, and second with respect to the most accurate solution of the TT-solver, i.e. for $\epsilon=10^{-6}$. The corresponding results are presented in Table \ref{tab:simple_gene_epsilon}, where the simulation time and memory requirement for storing the joint in the TT-format are also reported.
As can be seen from the table, the accuracy of the MCMC solution is reached already for a TT-sover accuracy of $\epsilon=10^{-4}$. 
Looking at the memory consumption and the execution time, they both increase for a higher TT-solver accuracy, which is expected since more solver iterations are needed to reach the desired residual.

\begin{table}[h]
	\caption{Simple gene expression model error analysis with respect to solver accuracy $\epsilon$.}
	\label{tab:simple_gene_epsilon}
	\begin{ruledtabular}
		\begin{tabular}{lllll}
			$\epsilon$ & \multicolumn{1}{p{1.5cm}}{\centering Error w.r.t.\\  MCMC}  &\multicolumn{1}{p{1.5cm}}{\centering Error w.r.t.\\  $\epsilon=10^{-6}$}  & Time [min] & Memory [MB] \\
			\hline
			$10^{-3}$ & $7.35\times 10^{-3}$ & $4.376\times 10^{-3}$  & $2.4$ & $1.8$  \\ 
			$10^{-4}$ & $2.35\times 10^{-3}$ & $1.309\times 10^{-3}$  & $7.4$ & $4.57$  \\ 
			$10^{-5}$ & $3.59\times 10^{-3}$ & $6.399\times 10^{-5}$  & $21$ & $9.29$ \\ 
			$10^{-6}$ & $3.64\times 10^{-3}$ & -  & $60$ & $17.74$ \\ 
		\end{tabular}
	\end{ruledtabular}
\end{table}

Additionally, the dimension of the parameter basis has been investigated. If the tensor product basis is constructed using univariate B-spline bases of dimension $16$, the prior can be well approximated. However during the inference, the decrease of the variance of the joint leads to an incapability to resolve the posterior, since a finer basis is needed. If the discretization is increased to 32 for every parameter, the oscillations become negligible.

As a conclusion, the limiting factor in the inference framework seems to be the accuracy of the TT-solver. For the purpose of inference, however, a relative residual value of $\epsilon=10^{-5}$ seems to be sufficient for obtaining an accurate approximation and an acceptable computational time.

\subsubsection{Gene expression model with feedback}

\begin{figure}[ht!]
	\centering
	\includegraphics[width=0.5\textwidth]{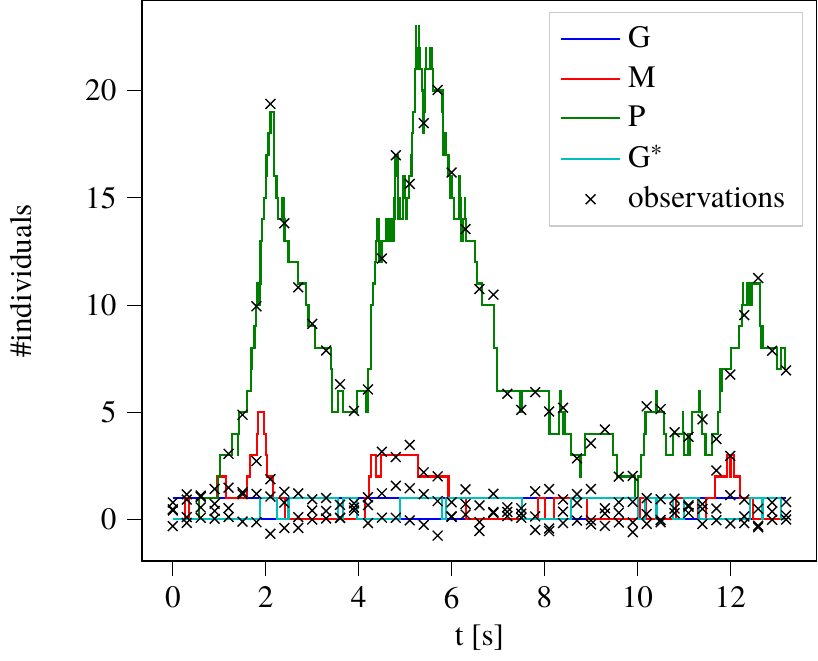}
	\caption{Noisy observation sample for the 3 stage gene expression model (number of observations is 45).  }
	\label{fig:3stage_sample}
\end{figure}

The second model where the parameter identification is employed is the 3-stage gene expression model with feedback loop\cite{_cal_2019}. The reactions as well as the reaction rates are presented in Table \ref{tab:gene}. A realization is drawn using the SSA and equidistant sampling is performed with additive Gaussian noise (see Fig. \ref{fig:3stage_sample}).

\begin{table}
	\caption{Reactions of the 3 stage gene expression model.}
	\label{tab:gene}
	\begin{ruledtabular}
		\begin{tabular}{cll}
			Reaction & $\alpha_m(\bm{x})$ & Rate $c_i$\\
			\hline
			$G \rightarrow G+M$ & $c_1x_1$  & $4.0$   \\ 
			$M \rightarrow M+P$ & $c_2x_2$  & $10.0$ \\ 
			$M \rightarrow \emptyset$ & $c_3x_2$  & $1.0$ \\ 
			$G+P \rightarrow G^*$ & $c_4x_1x_3$  & $0.2$  \\ 
			$G^* \rightarrow G+P$ & $c_5x_4$  & $0.6$   \\ 
			$P \rightarrow \emptyset$ & $c_6x_3$  & $1.0$  \\ 
		\end{tabular}
	\end{ruledtabular}
\end{table}

\begin{figure*}[ht!]
	\centering
	\includegraphics[width=1\textwidth]{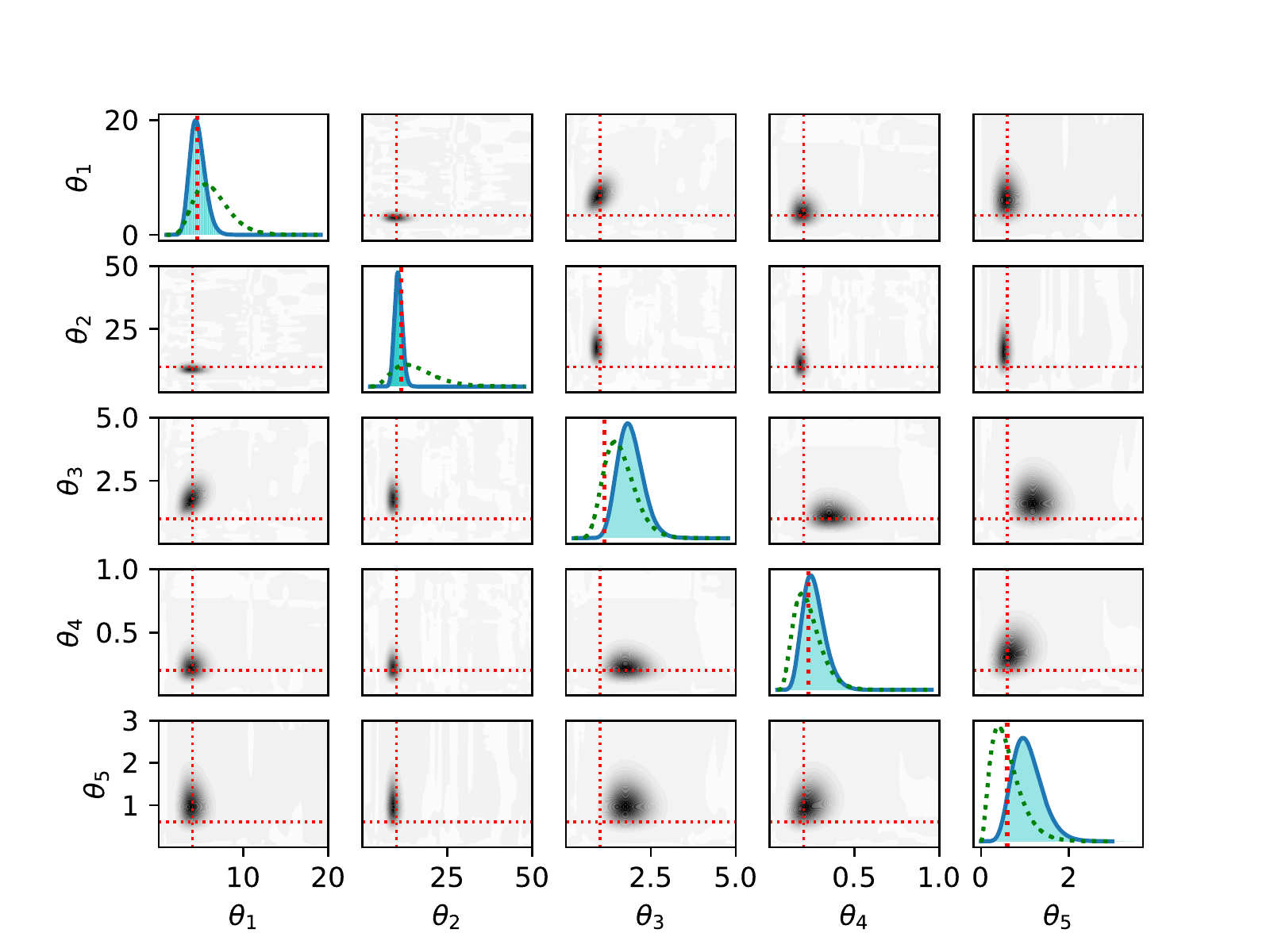}
	\caption{Posterior marginal distributions for the five unknown reaction rates of the 3 stage gene expression model. The black regions correspond to high density of the PDF. The exact parameters are marked with the red dashed lines. For the 1D marginals, a histogram of the posterior sample is represented as a reference, as well as the prior (green dashed lines). }
	\label{fig:3stage_posterior}
\end{figure*}

The parameters to be identified are in this case $\bm{\theta}=(c_1,...,c_5)$ and the parameter space is bounded to $[0,5c_i]$. For the priors, we choose again independent Gamma distributions, which are truncated within the parameter space. The parameter dependence is approximated using a tensor product basis of univariate quadratic B-splines with dimension $64$. The tolerance of the TT-solver is set to $10^{-5}$ in terms of relative residual.

The results are reported in Fig. \ref{fig:3stage_posterior} where we can see the visual match between the histograms and the 1D marginals on the diagonal. The expected value and variance of the posterior are:
\begin{align*}
\mathbb{E}[\bm{\theta}] &= (4.0358, 9.1720, 1.8398, 0.2378, 1.0686),\\
\mathbb{V}[\bm{\theta}] &= (0.9649, 1.5117, 0.1669, 0.005187, 0.1172).
\end{align*}
As a comparison, the mean and variance of the reference posterior sample are computed using MCMC:
\begin{align*}
\bm{\mu}_{\bm{\theta}} &= (4.0503, 9.1995, 1.8443, 0.2379, 1.0680),\\
\bm{\sigma}_{\bm{\theta}}^2 &= (0.9874, 1.2367, 0.4123, 0.0720, 0.3467).
\end{align*}
The relative error between the reference and the TT-solver-based modes is in the range of $10^{-3}$ for the expectation and $10^{-2}$ for the variance. The limiting factor is in this case the small MCMC sample size. 
Using the TT solver, the execution time for this test case is $50$ minutes. Regarding storage needs, only $\approx 12$ MB of RAM are used. As a comparison, the MCMC simulation took approximately $2.5$ days to complete for a sample size equal to $5\cdot 10^{5}$.

\subsubsection{SEIQR model}
\begin{figure}[H]
	\centering
	\includegraphics{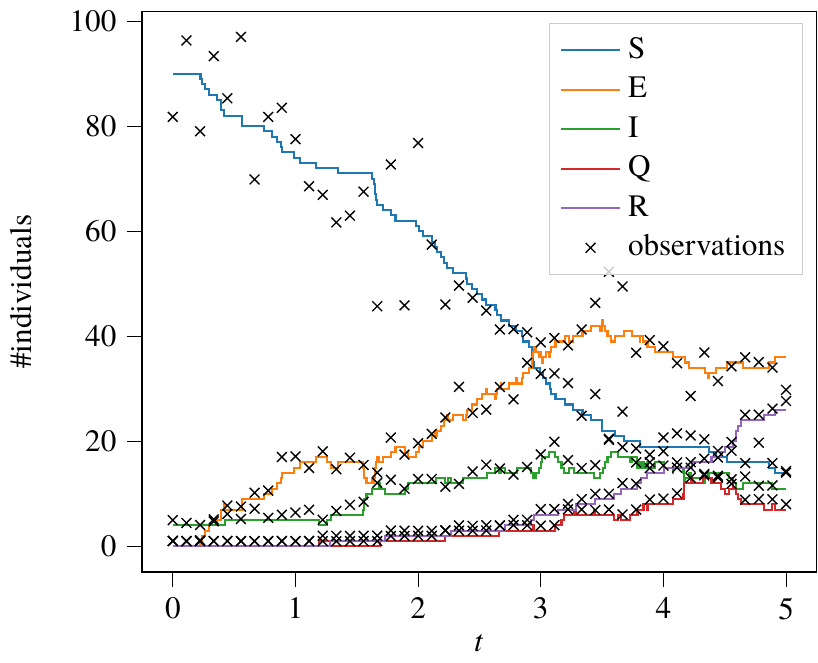}
	\caption{Noisy observation sample for the SEIQR model (number of observations is 45).  }
	\label{fig:seiqr_sample}
\end{figure}

The model considered now is an extension of the SEIR model presented in the filtering section and has one additional species: quarantined (Q). The modified reactions are found in Table \ref{tab:seiqr}. We infer in this case the parameters $\bm{\theta}=(c_1,c_2,c_3,c_4)$ from $45$ observations affected by lognormal noise (see Fig. \ref{fig:seiqr_sample}). The species Susceptibe and Exposed are observed with a higher degree of uncertainty while Quarantined and Recovered are observed exactly. The execution time for a TT-solver accuracy of $\epsilon=10^{-5}$ is $\approx 55$ minutes with a maximum posterior size in the QTT-format of $\approx30$ MB. As a comparison, the chosen state truncation of $(128,64,64,32,32)$ would require $\approx 4.2$ GB only for storing the state for one parameter realization. The storage complexity for the parameter-dependent CME operator in the QTT format is $\approx 200$ KB.

\begin{figure*}[ht!]
	\centering
	\includegraphics{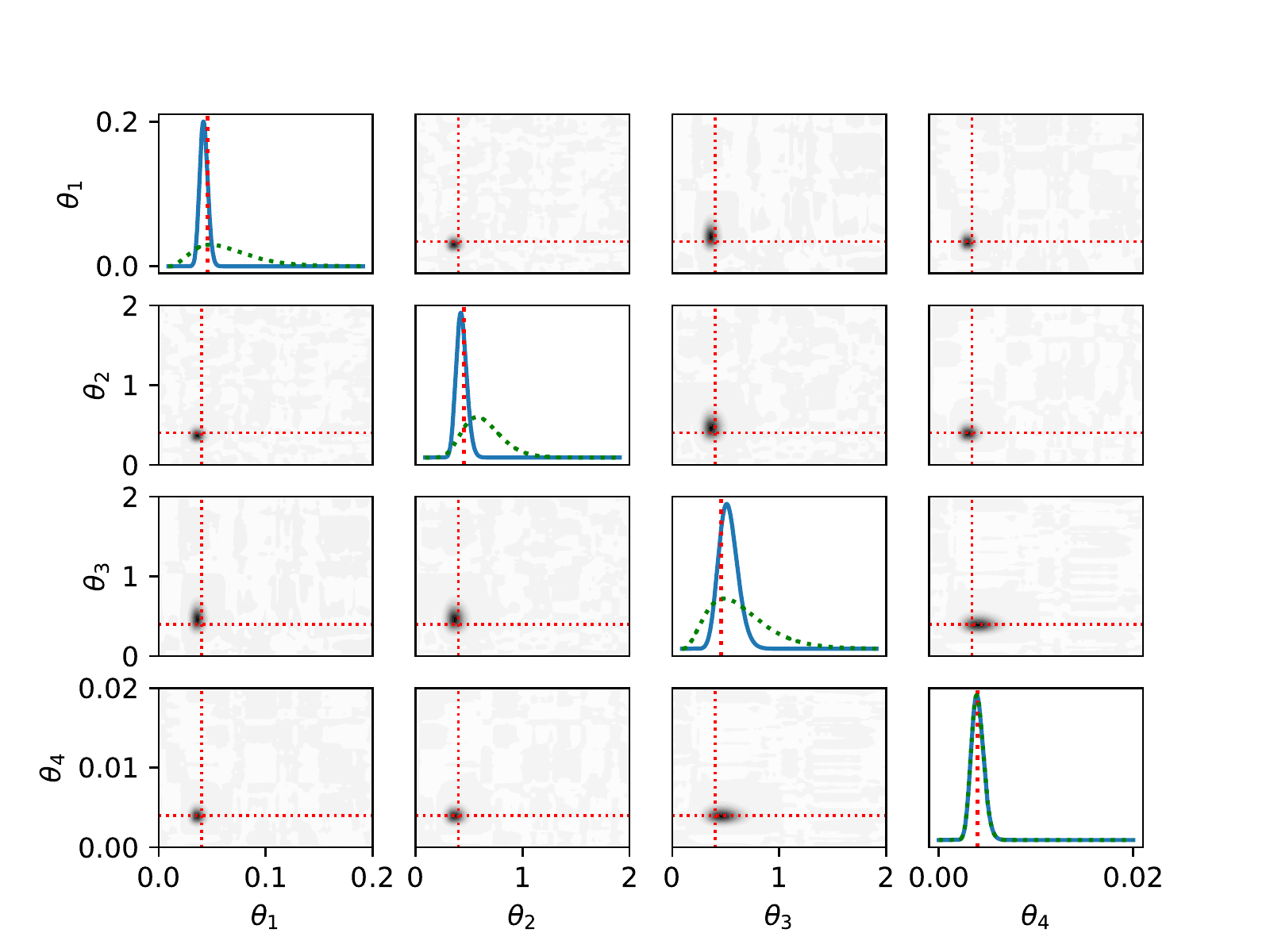}
	\caption{Posterior marginal distributions for the four unknown reaction rates of the SEIQR model. The black regions correspond to high density of the PDF. The exact parameters are marked with the red dashed lines and the prior with green dashed lines. }
	\label{fig:seiqr_posterior}
\end{figure*}

For this setup, the variance of the approximated posterior (see Fig. \ref{fig:seiqr_posterior}) is two orders lower than the prior for the first parameter and one order lower for the second and third parameters. This implies a higher confidence in the reconstruction of the parameters, which also indicates the need for a denser basis. In this case, choosing less than 64 points per parameter would lead to oscillations and inability to infer the posterior. This can be overcome by adaptively reducing the bounds of the parameter domain and re-interpolating the posterior on the new basis.

\begin{table}
	\caption{Reactions of the SEIQR model.}
	\label{tab:seiqr}
	\begin{ruledtabular}		\begin{tabular}{clll}
			Reaction & $\alpha_m(\bm{x})$ & Rate $c_i$ & Description\\
			\hline
			$S+I \rightarrow E+I$ & $c_1x_1x_3$  & $0.04$ & Susceptible becomes exposed \\ 
			$E \rightarrow I$ & $c_2x_2$  & $0.4$ & Exposed becomes infected \\ 
			$I \rightarrow Q$ & $c_3x_3$  & $0.4$ & Infected is quarantined \\
			$I \rightarrow \emptyset $ & $c_4x_3$  & $0.004$ & Infected individual dies \\ 
			$I \rightarrow R$ & $c_5x_3$  & $0.12$ & Infected recovers with immunity \\
			$Q \rightarrow R$ & $c_6x_4$  & $0.8765$ & Quarantined recovers with immunity \\
			$I \rightarrow S$ & $c_7x_3$  & $0.01$ & Infected recovers without immunity \\
			$Q \rightarrow S$ & $c_8x_4$  & $0.01$ & Quarantined recovers without immunity \\
			$\emptyset \rightarrow S$ & $c_9$  & $0.01$ & New susceptible individual\\			
\end{tabular}	\end{ruledtabular}\end{table} 

\section{Conclusion}
\label{sec:conclusion}
We presented a method based on the TT decomposition to solve the CME, either in its standard form or including parameter dependencies, and approximate the joint distribution over the state-parameter space, including the time dependency as well.
Using the considered TT-framework, inference tasks such as state smoothing and parameter identification can be performed accurately and efficiently.
The proposed TT-framework is also applied to solve the inverse problem of identifying the values of the parameters governing the system under investigation from noisy state observations. 
The resulting numerical approximation of the posterior PDF over the truncated parameter space can be efficiently stored and manipulated in the TT format. 

A series of numerical experiments with a simple gene expression model and with the SEIR virus model show clearly that the state-time TT-approximation reduces the storage needs of the CME to a mere fraction of what a standard CME solution method requires.
Moreover, by performing multilinear algebraic operations in the TT-format, the execution time is significantly reduced as well.
Similar benefits are observable in the context of inference tasks, where the proposed TT-based filtering, smoothing, and parameter identification approaches yield accurate results for a significantly reduced computational cost.

While standard inference procedures for the single trajectory setting such as MCMC require repeated solutions of the CME for different parameter configurations, the joint approach presented here requires only one forward pass on the augmented state space. One drawback of the parameter space discretization is that it can cause problems when the posterior is much more concentrated then the prior. As demonstrated on the SEIQR model, this can be overcome by dynamically adapting the basis. 

In this work, we have focused on inferring the rate constants of structurally known models from a single trajectory. An important direction for future research is to extend the approach to multiple trajectories with shared parameters. Another interesting direction is to consider different types of uncertainty, e.g. the involved species or the types of the reaction. Since the presented method is fully Bayesian, this could be realized by scoring different candidate structures via Bayesian model comparison.

\section*{Acknowledgment}
The work of I.G.~Ion is supported by the Graduate School Computational Engineering within the Centre for
Computational Engineering at Technische Universit\"at Darmstadt. D. Loukrezis is supported by the German Federal Ministry for Education and Research (BMBF) via the research contract 05K19RDB.

\section*{Data availability}

The data and code that support the findings of this study are freely available at  \emph{https://github.com/ion-g-ion/paper-cme-tt}.

\bibliography{main_jcp_review_2}

\end{document}